%% file: B2G-15-008_temp.tex
\pdfoutput=1

\documentclass[11pt,twoside,a4paper,cmspaper,final,collab]{cms-tdr}
\def\svnVersion{406224}\def\cmsCernNoTag{CERN-EP/2016-279}\def\cmsCernDate{\today}\def\cmsMessage{Published in Physics Letters B as \href{http://dx.doi.org/10.1016/j.physletb.2017.05.019}{\doi{10.1016/j.physletb.2017.05.019.}}}

\begin{document}\cmsNoteHeader{B2G-15-008}

\hyphenation{had-ron-i-za-tion}
\hyphenation{cal-or-i-me-ter}
\hyphenation{de-vices}
\RCS$Revision: 399450 $
\RCS$HeadURL: svn+ssh://svn.cern.ch/reps/tdr2/papers/B2G-15-008/trunk/B2G-15-008.tex $
\RCS$Id: B2G-15-008.tex 399450 2017-04-17 21:01:36Z alverson $
\newlength\cmsFigWidth
\ifthenelse{\boolean{cms@external}}{\setlength\cmsFigWidth{0.85\columnwidth}}{\setlength\cmsFigWidth{0.4\textwidth}}
\ifthenelse{\boolean{cms@external}}{\providecommand{\cmsLeft}{top\xspace}}{\providecommand{\cmsLeft}{left\xspace}}
\ifthenelse{\boolean{cms@external}}{\providecommand{\cmsRight}{bottom\xspace}}{\providecommand{\cmsRight}{right\xspace}}
\providecommand{\cPQT}{\ensuremath{\cmsSymbolFace{T}}\xspace}
\providecommand{\cPaQT}{\ensuremath{\overline{\cmsSymbolFace{T}}}\xspace}
\providecommand{\twothird}{\ensuremath{2\hspace{-0.06em}/\hspace{-0.08em}3}\xspace}
\providecommand{\bW}{\ensuremath{\PQb\PW}\xspace}
\providecommand{\tZ}{\ensuremath{\PQt\cPZ}\xspace}
\providecommand{\tH}{\ensuremath{\PQt\PH}\xspace}
\newcommand{\AMCATNLO} {a\textsc{mc@nlo}\xspace}
\newcommand{\amcatnlo}   {\MADGRAPH5\_a\MCATNLO\xspace}
\providecommand{\MT}{\ensuremath{M_{\cPQT}}\xspace}
\cmsNoteHeader{B2G-15-008}
\title{Search for single production of a heavy vector-like T quark decaying to a Higgs boson and a top quark with a lepton and jets in the final state}

\date{\today}

\abstract{
A search for single production of vector-like top quark partners (T)
decaying into a Higgs boson and a top quark is performed
using data from pp collisions at a centre-of-mass energy of 13\TeV
collected by the CMS experiment at the CERN LHC, corresponding to an integrated
luminosity of 2.3\fbinv. The top quark decay includes an electron
or a muon while the Higgs boson decays into a pair of b quarks. No
significant excess over standard model backgrounds is
observed. Exclusion limits on the product of the production cross section
and the branching fraction are derived in the T quark mass range 700 to 1800\GeV. For a
mass of 1000\GeV, values of the product of the production cross section and the branching
fraction greater than 0.8 and 0.7\unit{pb} are excluded at 95\% confidence level, assuming
left- and right-handed coupling of the T quark to standard model
particles, respectively.
This is the first analysis setting exclusion limits on the cross section of singly
produced vector-like T quarks at a centre-of-mass energy of 13\TeV.
}

\hypersetup{%
pdfauthor={CMS Collaboration},%
pdftitle={Search for single production of a heavy vector-like T quark decaying to a Higgs boson and a top quark with a lepton and jets in the final state},%
pdfsubject={CMS},%
pdfkeywords={CMS, B2G, vector-like-quark, VLQ, Tprime, Higgs tagging, physics}}

\maketitle

\section{Introduction}

Over the past decades several theoretical models have been formulated trying to give new insights into electroweak symmetry breaking and the mechanisms that stabilise the mass of the Higgs boson.
Many of these models predict the existence of heavy vector-like quarks. Examples are little Higgs models~\cite{ArkaniHamed:2001nc,Schmaltz:2005ky,Perelstein:2003wd}, models with extra dimensions~\cite{Antoniadis:2001cv,Hosotani:2004wv}, and composite Higgs boson models~\cite{Dugan:1984hq,Kaplan:1991dc,Agashe:2004rs,Contino:2006qr,DeSimone2013}.

The distinctive property of vector-like quarks is that their left- and right-handed components transform in the same way under the  electroweak symmetry group $\mathrm{ SU(2)_L \times U(1)_Y }$ of the standard model (SM).
As a consequence, vector-like quarks can obtain mass through direct mass terms in the Lagrangian of the form $m \overline{\psi}\psi$, unlike the SM chiral quarks, which obtain mass through Yukawa coupling.

The discovery of a Higgs boson by the ATLAS~\cite{Aad:2012tfa} and CMS~\cite{Chatrchyan:1471016,Chatrchyan:2013lba} Collaborations and the electroweak fits within the framework of the SM~\cite{PhysRevD.86.013011} strongly disfavour the existence of a fourth generation of chiral fermions.
Given the limited impact that vector-like quarks have on the properties of the SM Higgs boson, they are not similarly constrained~\cite{PhysRevD.88.094010}.

This letter presents the results of the first search for singly produced vector-like top quark partners with charge $+\twothird$ (\cPQT) at a centre-of-mass energy of $\sqrt{s}=13\TeV$.
Single production is of particular interest, since its rate dominates over the rate of pair production at large quark masses.
Many of the models mentioned above predict that the \cPQT quark will predominantly decay to third-generation SM quarks via three channels: \tH, \tZ, and \bW~\cite{PhysRevD.88.094010}.
Searches for \cPQT quarks have been performed by the ATLAS and CMS Collaborations setting lower limits on the \cPQT quark mass ranging from 715 to 950\GeV for various \cPQT quark branching fractions~\cite{Aad:2014efa,Aad:2015gdg,Aad:2015kqa,Aad:2015voa,Aad:2016qpo,Khachatryan:2015oba,B2G14002paper}.

\begin{figure}[htb]
\centering
\includegraphics[width=0.4\textwidth]{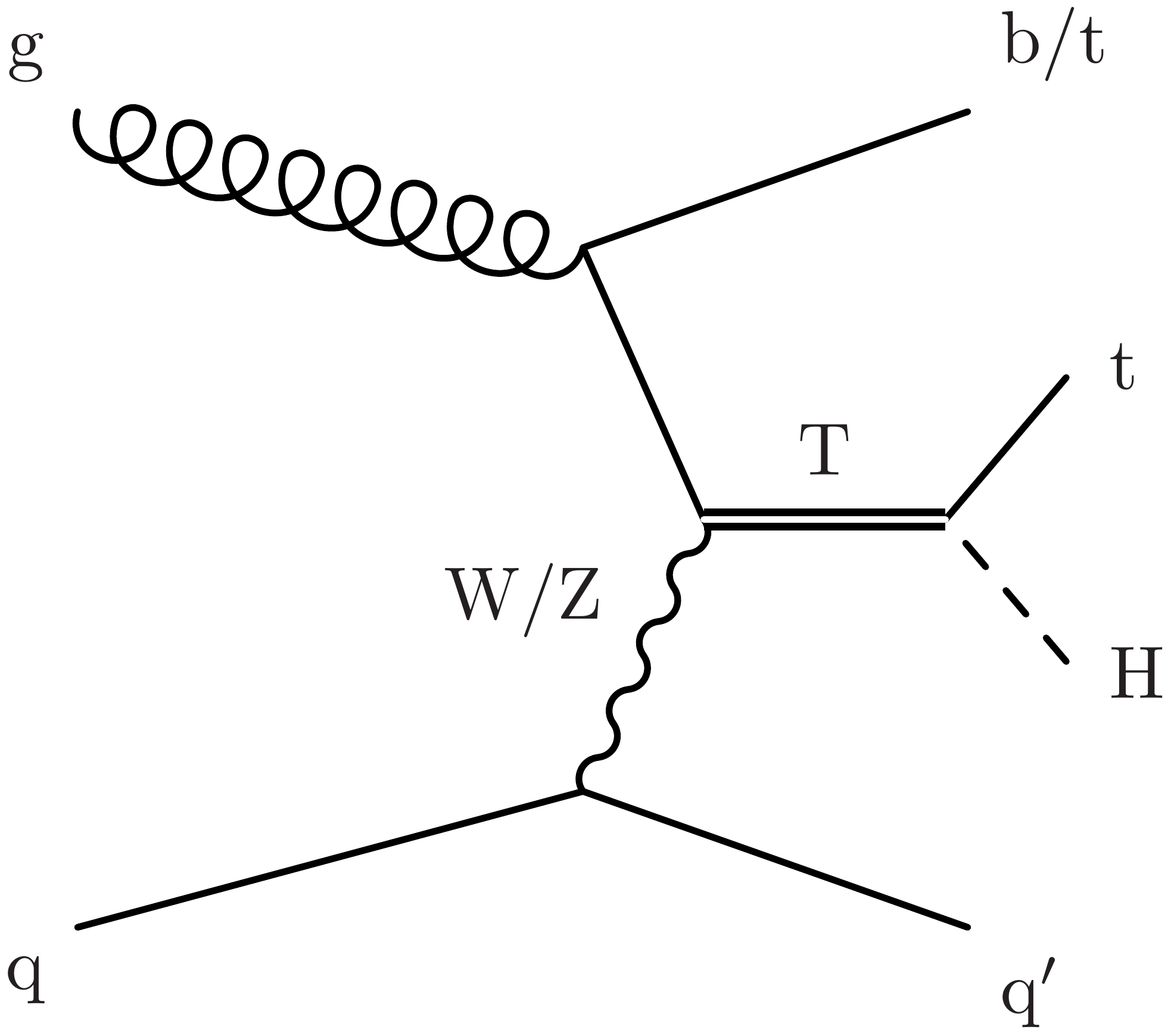}
\caption{
Feynman diagram of the production and decay mechanisms of a vector-like \cPQT quark, as targeted in this analysis.
}
\label{fig_vlq_feynman}
\end{figure}

While most of the past searches considered pair production of the \cPQT quarks via the strong interaction, the single production mode where the \cPQT quark is produced via the weak interaction has recently been investigated by the ATLAS Collaboration~\cite{Aad:2014efa,Aad:2015voa,Aad:2016qpo} at 8\TeV, and is targeted in this letter.
The strength of the \cPQT quark coupling to electroweak bosons has an effect both on the cross section and the width of the \cPQT quark~\cite{Matsedonskyi2014}.
There are no a priori constraints on the electroweak \cPQT quark coupling.
Therefore, not only the general coupling to the electroweak sector but the couplings of the \cPQT quark to \bW, \tZ, and \tH can also take arbitrary values.
The present analysis targets decays of the \cPQT quark into a Higgs boson and a top quark.
It will be sensitive to the existence of a \cPQT quark only if sufficiently large couplings to \bW or \tZ are present as well, since the \cPQT quark production through a Higgs boson is strongly suppressed.
An example of a Feynman diagram for this process is shown in Fig.~\ref{fig_vlq_feynman}.

The analysis is performed on the proton-proton collision data collected during 2015 by the CMS experiment at the CERN LHC at  $\sqrt{s}=13\TeV$.
The search is optimised for decays of the \cPQT quark into a Higgs boson and a top quark, where the top quark decay includes a lepton (electron or  muon) and the Higgs boson is required to decay into \PQb quarks.
For a \cPQT quark mass in the \TeV range, the Higgs boson and the top quark obtain large Lorentz boosts leading to merged jets and nonisolated leptons in the final state.
Jet substructure analysis in combination with algorithms for the identification of \PQb quark jets (\PQb tagging) can efficiently identify boosted decays of the Higgs boson into \PQb quark pairs \cite{B2G14002paper}.
An additional distinctive feature of the signal is the presence of a jet in the regions close to the beam pipe, a so-called forward jet.
This jet results from the light-flavour quark that is produced in association with the \cPQT quark.
Background processes due to top quark pair production are dominant, followed by \PW+jets and quantum chromodynamics (QCD) multijet processes.

For every event, a \cPQT quark candidate four-momentum is reconstructed, with mass \MT.
Events are selected by imposing requirements on the \cPQT quark candidate and other attributes of the event.
The \MT variable is used as the final discriminant in a combined signal plus background fit to the data.
The shape of the total background is estimated from a signal-depleted region in the recorded data.

\section{The CMS detector}
\label{detector}

The central feature of the CMS apparatus is a superconducting solenoid of 6\unit{m} internal dia\-meter, providing a magnetic field of 3.8\unit{T}. Within the solenoid volume are a silicon pixel and strip tracker, a lead tungstate crystal electromagnetic calorimeter (ECAL), and a brass and scintillator hadron calorimeter (HCAL), each composed of a barrel and two endcap sections. Forward calorimeters extend the coverage provided by the barrel and endcap detectors to regions close to the beam pipe. Muons are measured in gas-ionisation detectors embedded in the steel flux-return yoke outside the solenoid.
A more detailed description of the CMS detector, together with a definition of the coordinate system used and the relevant kinematic variables, can be found in Ref.~\cite{Chatrchyan:2008zzk}.

A particle-flow (PF) algorithm~\cite{CMS-PAS-PFT-09-001,CMS-PAS-PFT-10-001} is used to combine information from all CMS subdetectors in order to reconstruct and identify individual particles in the event: photons, electrons, muons, and charged and neutral hadrons.
The energy of photons is directly obtained from the ECAL measurement.
The energy of electrons is determined from a combination of the electron momentum at the primary interaction vertex  determined by the tracker, the energy of the corresponding ECAL cluster, and the energy sum of all bremsstrahlung photons spatially compatible with originating from the electron track.
The momentum resolution for electrons with transverse momentum $\pt \approx 45\GeV$ and above from $\cPZ \to \Pe \Pe$ decays ranges from 1.7\% for non-showering electrons in the barrel region to 4.5\% for showering electrons in the endcaps~\cite{Khachatryan:2015hwa}.
Muons are measured in the pseudorapidity range $\abs{\eta}< 2.4$ with detection planes made using three technologies: drift tubes, cathode strip chambers, and resistive-plate chambers.
Matching muons to tracks measured in the silicon tracker results in a relative \pt resolution of 1.2--2.0\% for muons with $20 <\pt < 100\GeV$  in the barrel and better than 6\% in the endcaps.
The \pt resolution in the barrel is better than 10\% for muons with \pt up to 1\TeV~\cite{Chatrchyan:2012xi}.
The energy of charged hadrons is determined from a combination of their momentum measured in the tracker and the matching of ECAL and HCAL energy deposits, corrected for the response function of the calorimeters to hadronic showers. Finally, the energy of neutral hadrons is obtained from the corresponding corrected ECAL and HCAL energy.

Jets are reconstructed from the individual particles identified by the PF event algorithm, clustered by the anti-$k_t$ algorithm~\cite{Cacciari:2008gp, Cacciari:2011ma}.
Two different jet sizes are used independently: jets with a size parameter of 0.4 (``AK4 jets'') and 0.8 (``AK8 jets'').
Jet momentum is determined as the vector sum of the charged particle momenta in the jet that are identified as originating from the primary interaction vertex, and the neutral particle momenta.
An area-based correction is applied to jet energies to take into account the contribution from additional proton-proton interactions within the same or adjacent bunch crossings (``pileup'')~\cite{Cacciari:2008gn}.
The energy of a jet is found from simulation to be within 5--10\% of the true jet momentum at particle level over the entire \pt spectrum and detector acceptance.
Jet energy corrections are derived from simulation, and are confirmed with in situ measurements of the energy balance in dijet and photon+jet events~\cite{Chatrchyan:2011ds}.
A smearing of the jet energy is applied to simulated events to mimic detector re\-so\-lu\-tion effects observed in data. For the identification of \PQb jets, the combined secondary vertex \PQb tagging algorithm is used~\cite{Chatrchyan:2012jua}. The algorithm  uses information from  secondary \PQb hadron decay vertices to distinguish \PQb jets from other jet flavours.
The jet energy resolution is typically 15\% at 10\GeV, 8\% at 100\GeV, and 4\% at 1\TeV. Jets are reconstructed up to $\abs{\eta} = 5$ while \PQb tagging is restricted by the tracker acceptance to $\abs{\eta} < 2.4$.

The missing transverse momentum vector \ptvecmiss is defined as the negative vector sum of the \pt of all PF particle candidates in an event. Its magnitude is referred to as \ETmiss.

\section{Data and simulated samples}
\label{samples}

Events in the electron channel are selected using an electron trigger, which requires an electron with $\pt>45$\GeV and the additional presence of at least two jets, with $\pt>200$\GeV and $50$\GeV, respectively for the jets with the highest and second highest \pt.
Events in the muon channel are collected with a single-muon trigger, requiring the presence of a muon candidate with $\pt>45$\GeV and $\abs{\eta}< 2.1$.
The muon trigger does not require a jet.
Neither of the triggers places any requirement on the isolation of the leptons.
If an event is selected by both the electron and the muon trigger, which happens almost exclusively in top quark pair events containing an electron and a muon, it is assigned to the muon channel.
The data collected with the muon trigger correspond to a luminosity of ${\cal L} = 2.3 \,$fb$^{-1}$, while the electron trigger provides a luminosity ${\cal L} = 2.2 \,$fb$^{-1}$.

Signal samples are generated using \amcatnlo~2.2.2~\cite{Alwall:2014hca} at leading order (LO) QCD accuracy.
The cross section to produce a heavy \cPQT quark decaying to top quark and Higgs boson in association with a bottom or top quark is set to 1\unit{pb} unless indicated differently.
Signal masses are simulated between 700 and 1800\GeV in steps of 100\GeV, assuming a fixed \cPQT quark width of 10\GeV.
This width corresponds to a narrow width approximation, meaning that the experimental resolution is much larger than the width used in generating the samples.
Generation of both the \cPQT quark and its antiquark are included, with the positive charge having a higher occurrence because of the larger density of positively charged u quarks in the proton.
Only the left- (right-) handed \cPQT quark chiralities in association with a bottom (top) quark are considered, as only those are allowed in the singlet (doublet) scenario of the simplest Simplified Model~\cite{Matsedonskyi2014}.
Left- and right-handed production of the \cPQT quark are simulated in separate samples.

Background events from top quark pair production and electroweak production of a single top quark in the tW-channel are simulated at next-to-leading order (NLO) with the \POWHEG~2.0 generator~\cite{Nason:2004rx,Frixione:2007vw,Alioli:2010xd,Frixione:2007nw}.
The \amcatnlo at NLO accuracy is used to generate samples of single top quarks in the s- and t-channels.
The generation of the \PW+jets and \cPZ+jets events is performed at LO with the \amcatnlo, with up to four partons included in the matrix element calculations, matched to parton showers using the so-called MLM scheme~\cite{mlm}.
All samples are interfaced with \PYTHIA~8.212~\cite{Sjostrand:2006za,Sjostrand:2014zea}, tune CUETP8M1~\cite{GEN-14-001} for the description of hadronisation and fragmentation.
The QCD multijet background events are generated with \PYTHIA for both matrix element and showering descriptions.

All samples are generated using NNPDF~3.0~\cite{Ball:2014uwa} parton distribution functions (PDFs) either at LO or at NLO, to match the precision of the matrix element calculation.
The effects of pileup are simulated in all samples by adding simulated minimum bias events to the hard scattering process, according to a distribution having an average multiplicity of 11 collisions per bunch crossing, as observed in data.

All events are processed through a full simulation of the CMS detector using  \textsc{Geant4}~\cite{Agostinelli2003250,1610988}.

\section{Event reconstruction and selection}
\label{sec_reco_sel}

Primary vertices are reconstructed using a deterministic annealing filtering algorithm~\cite{Chatrchyan:2014fea}.
The leading vertex of the event is defined as the one with the largest sum of squared \pt of associated tracks.
Its position is reconstructed using an adaptive vertex fit~\cite{AVFitter} and is required to be within 24\unit{cm} in the $z$ direction and 2\unit{cm} in the $x$-$y$ plane of the nominal interaction point.

Events are required to have at least one lepton.
For large \cPQT quark masses, the top quark from the $\cPQT\to\PQt\PH$ decay has a significant Lorentz boost causing its products to be approximately collinear.
Thus as the lepton is not isolated from the \PQb quark jet (``\PQb jet''), no conventional isolation requirement (i.e. requiring the energy deposited in a cone around the lepton to be small) is applied.
In order to suppress QCD multijet events with a lepton (electron or muon) contained within an AK4 jet, the selection criteria $\Delta R (\ell, j) > 0.4$ or
$\pt^\text{rel}(\ell, j) > 40$\GeV are applied, where $\ell$ indicates the lepton and $j$ indicates the AK4 jet with lowest angular separation from the lepton.
The angular distance is defined as $\Delta R = \sqrt{\smash[b]{ (\Delta \eta)^2 + (\Delta \phi)^2}}$, where $\Delta\phi$ ($\Delta\eta$) is the difference in azimuthal angle (pseudorapidity) between the AK4 jet and the lepton, and $\pt^\text{rel}$ is the projection of the three-momentum of the lepton onto a plane perpendicular to the jet axis.
In addition to this selection, electrons (muons) must have $\pt > 50$ (47) \GeV and $\abs{\eta} < 2.5\,(2.1)$, to fall within a region where the trigger efficiency is constant.
In the case of more than one reconstructed lepton in the given channel, only the lepton with the highest \pt is used in the evaluation of physics quantities needed for this analysis and shown in the plots below.
The lepton isolation and trigger selection efficiencies are measured in the data and simulation as a function of $\eta$ and \pt of the lepton and are found to agree within their uncertainties.

All AK4 jets are required to have $\pt>30$\GeV.
If a selected lepton is found within a cone of $\Delta R(j, \ell)<0.4$ around the jet axis, the lepton four-momentum is subtracted from the uncorrected jet four-momentum and all jet energy corrections are applied thereafter.
AK4 jets with $\abs{\eta}>2.4$ are defined as ``forward jets''.
An event must have at least two AK4 jets.
The leading (subleading) AK4 jet \pt is required to exceed 250\,(70)\GeV in the electron channel and 100\,(50)\GeV in the muon channel.
The different \pt thresholds for the two channels are due to the tighter criteria of the electron trigger, which selects events with two high-\pt jets (Section~\ref{samples}).

Since the decay of a heavy \cPQT quark would produce high-energy final-state particles, all events are required to have $S_\mathrm{T}>400$\GeV, where $S_\mathrm{T}$ is defined as the scalar sum over \ETmiss, the \pt of the lepton and the transverse momenta of all selected AK4 jets in the event.

The AK8 jets are required to have $\pt>200$\GeV and $\abs{\eta}<2.4$.
The modified mass drop tagger algorithm~\cite{Dasgupta2013}, also known as the ``soft-drop'' algorithm with angular exponent $\beta = 0$, soft threshold $z_\text{cut} < 0.1$, and characteristic radius $R_0 = 0.8$~\cite{Larkoski:2014wba}, is used to remove soft, wide-angle radiation from the jet.
Subjets of AK8 jets are identified in the last reclustering step of the soft-drop algorithm.
The soft-drop jet mass scale and resolution have been estimated using a \ttbar control region.
This control region is defined by the baseline selection (see below) and additionally requiring two b-tagged AK4 jets as well as the N-subjettiness ratio $\tau_2/\tau_1$ to be smaller than 0.4~\cite{Thaler:2011gf, Thaler:2010tr} for the Higgs boson candidate (see below).
The mass scale is found to be compatible between data and simulation within uncertainties.
A degradation of the jet mass resolution of 10\% is applied in the simulation to match the resolution found in the data.

For the identification of \PQb jets, the combined secondary vertex \PQb tagging algorithm is used.
In this analysis, it is only applied to the final two soft-drop subjets of AK8 jets.
A working point that typically yields \PQb tagging efficiencies of approximately 80\% and misidentification rates from light-flavour jets of about 10\% in \ttbar events~\cite{Chatrchyan:2012jua} is chosen.
The \PQb tagging of subjets results in a better performance compared to the \PQb tagging of AK4 jets in \ttbar events, reducing the misidentification rate at the working point by a factor of approximately two.

In order to identify decays of the boosted Higgs boson to \PQb quark pairs (\PH tagging)~\cite{B2G14002paper}, the soft-drop mass of the jet, $M_{\PH}$, is required to be within $90<M_{\PH}<160\GeV$.
At least one Higgs boson candidate is required to be present and to have an angular separation of $\Delta R({\PH}, \ell)>1.0$ from the lepton.
The number of \PQb tagged subjets of the Higgs boson candidate is used to define the signal and background control regions.

To reconstruct the top quark, its decay into a bottom quark and a \PW~boson, with the \PW~boson subsequently decaying into a muon or electron and a neutrino, is assumed.
Using the $x$ and $y$ components of \ptvecmiss, the lepton four-momentum,
and the nominal mass of the \PW~boson (80.4\GeV), \cite{PDG:Agashe:2014kda} the $z$ component of the neutrino momentum is reconstructed by solving a qua\-drat\-ic equation, resulting in up to two solutions.
If a complex solution is obtained, only the real part is used.
Combining the four-momenta of these neutrino hypotheses and the lepton, up to two \PW~boson candidates are obtained.
Each \PW~boson candidate is paired to every central AK4 jet in the event, giving a number of reconstruction hypotheses for the top quark.
In order to accommodate final-state radiation from the top quark, further top quark reconstruction hypotheses are found by the addition of one more AK4 jet, such that one top quark candidate is established for every single AK4 jet and every possible combination of two AK4 jets.
The \PQb tagging information is not used in the top quark reconstruction.

Top quark and Higgs boson candidates are combined into pairs.
Combinations are rejected if any AK4 jet ($j_{\PQt}$) of the top quark candidate overlaps with the Higgs boson candidate within $\Delta R(j_{\PQt}, {\PH}) < 1.0$. This requirement ensures that there is no overlap or double counting of jets from the two jet collections with jet sizes 0.4 and 0.8.
The pair of candidates yielding the smallest $\chi^2$ value is used in the following analysis, where the $\chi^2$ function is defined as follows:
\ifthenelse{\boolean{cms@external}}{
\begin{multline*}\chi^2 = \left(\frac{M_{\PH, \mathrm{MC}} - M_{\PH}}{\sigma_{M_\PH, \mathrm{MC}}}\right)^2
         + \left(\frac{M_{\PQt, \mathrm{MC}} - M_{\PQt}}{\sigma_{M_{\PQt}, \mathrm{MC}}}\right)^2\\
         + \left(\frac{\Delta R({\PQt, \PH})_\mathrm{MC} - \Delta R({\PQt,\PH})}{\sigma_{\Delta R, \mathrm{MC}}}\right)^2 .
\end{multline*}
}{
\begin{equation*}\chi^2 = \left(\frac{M_{\PH, \mathrm{MC}} - M_{\PH}}{\sigma_{M_\PH, \mathrm{MC}}}\right)^2
         + \left(\frac{M_{\PQt, \mathrm{MC}} - M_{\PQt}}{\sigma_{M_{\PQt}, \mathrm{MC}}}\right)^2
         + \left(\frac{\Delta R({\PQt, \PH})_\mathrm{MC} - \Delta R({\PQt,\PH})}{\sigma_{\Delta R, \mathrm{MC}}}\right)^2 .
\end{equation*}
}
Here, $M$ denotes the mass of a candidate, and the \PH and \PQt subscripts stand for the Higgs boson and top quark candidates, respectively.
The ``MC'' subscript denotes that a quantity is derived from the signal simulation, using the correct pairing of the reconstructed objects based on Monte Carlo information.
Other quantities are obtained from the pair of top quark and Higgs boson candidates.

After event reconstruction, the selection is further refined by requiring a large separation of $\Delta R({\PQt, \PH})>2.0$ between the top quark and Higgs boson candidates. The top quark candidate must have $\pt>100\GeV$.

\begin{figure*}[htb]
\centering
\includegraphics[width=0.49\textwidth]{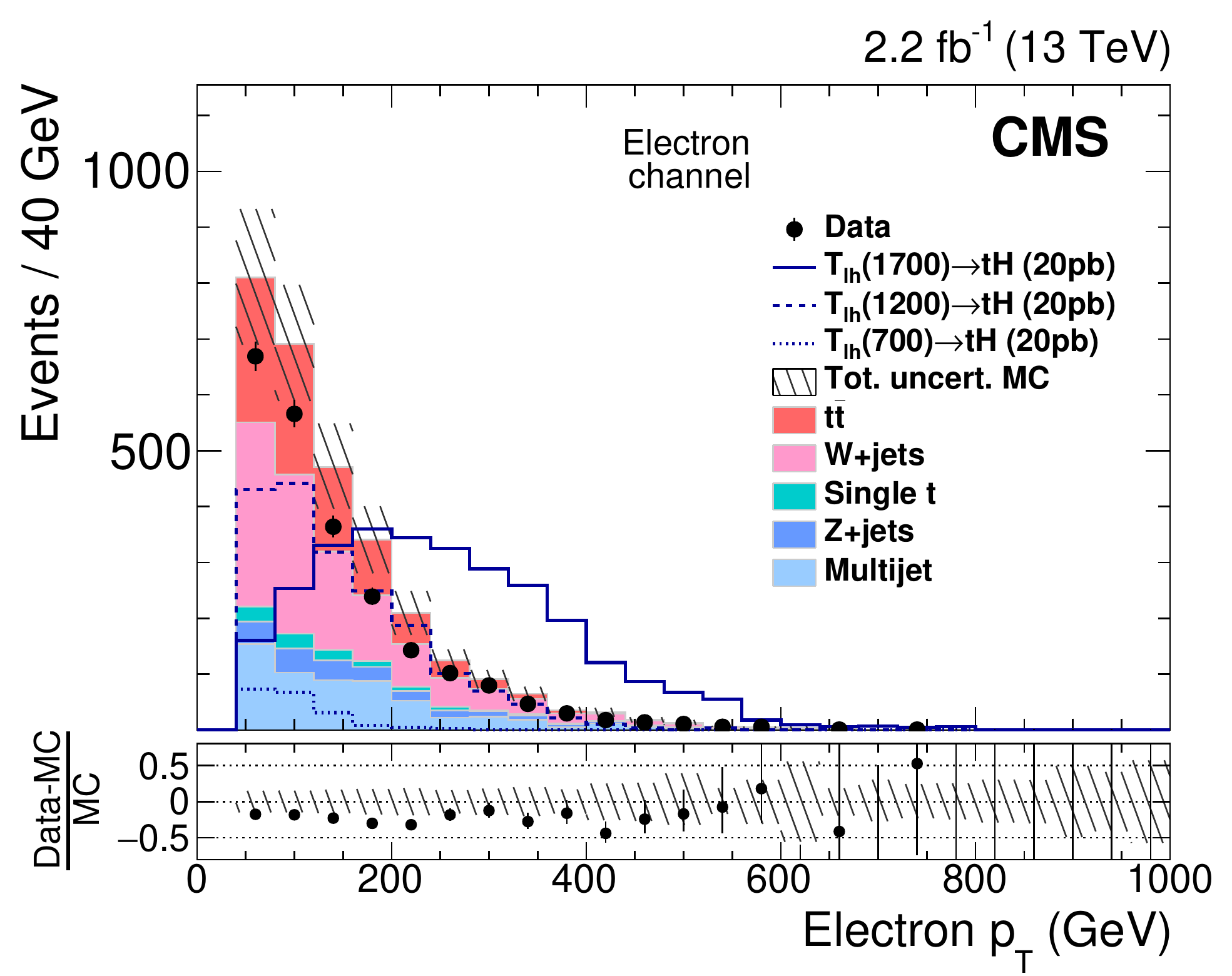}
\includegraphics[width=0.49\textwidth]{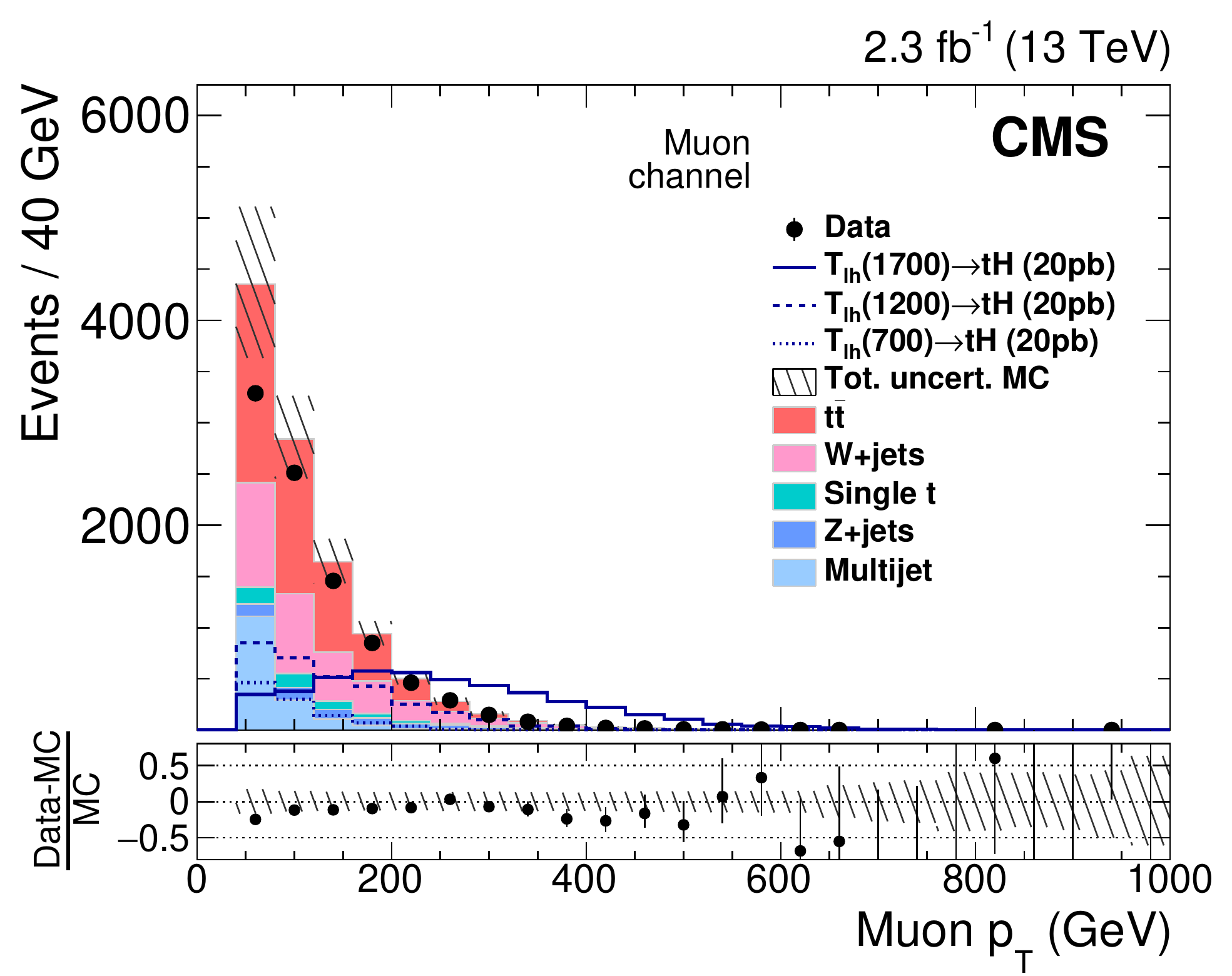}
\includegraphics[width=0.49\textwidth]{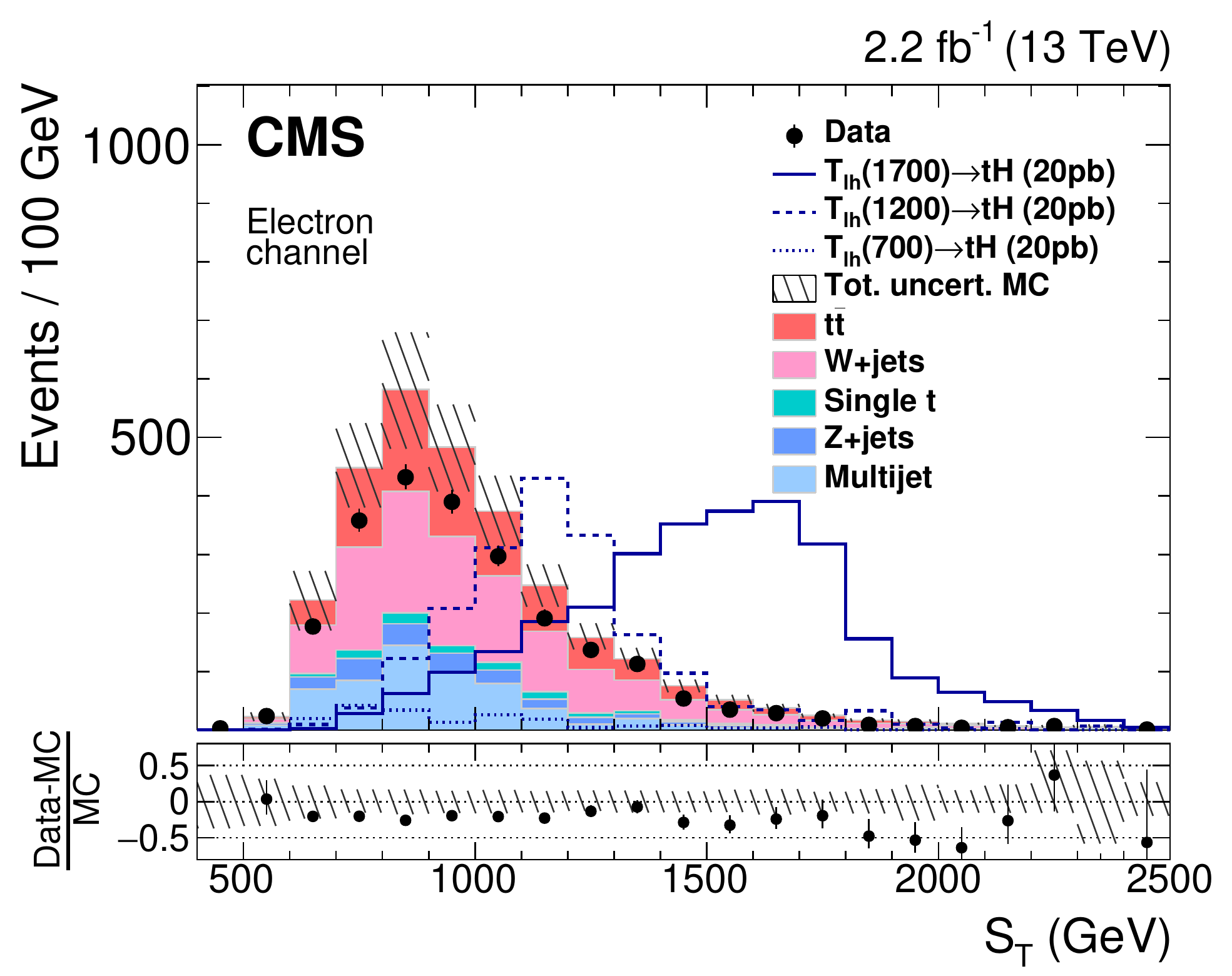}
\includegraphics[width=0.49\textwidth]{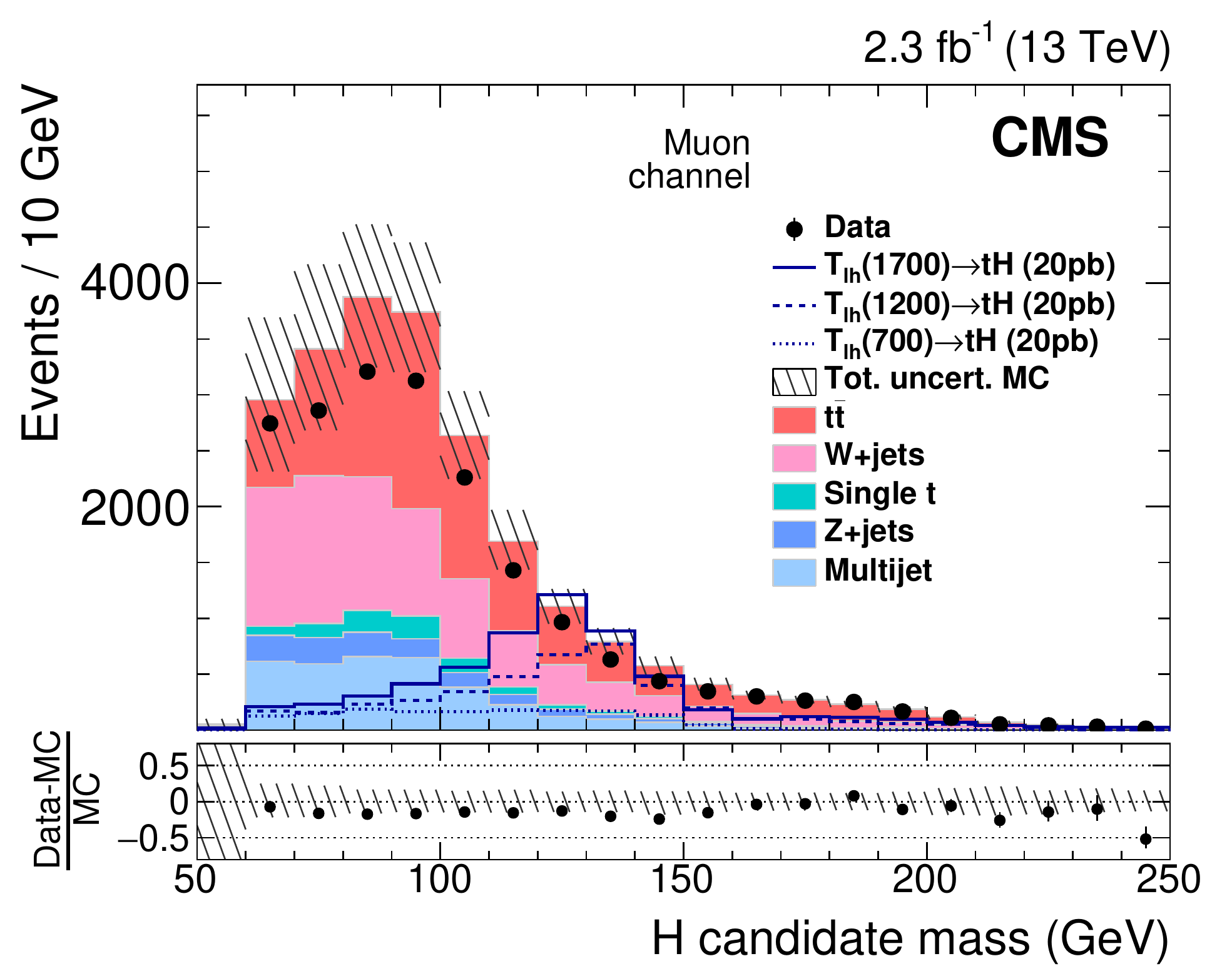}
\caption{
Distributions of kinematic variables after baseline selection.
Electron and muon \pt distributions are depicted in the upper-left and upper-right panels.
The lower-left panel shows $S_\mathrm{T}$ in the electron channel while the soft-drop mass of the Higgs boson candidate in the muon channel is depicted in the lower right.
The different background contributions are shown using full histograms while the open histograms are signal yields and the data are shown as solid circles.
The hatched bands represent the statistical and systematic uncertainties of the simulated event samples.
The systematic uncertainties include those discussed in Section~\ref{sec_uncert}, except the forward jet uncertainty.
Signal cross sections are enhanced to 20\unit{pb}.
}
\label{fig_obj}
\end{figure*}

\begin{figure*}[htb]
\centering
\includegraphics[width=0.49\textwidth]{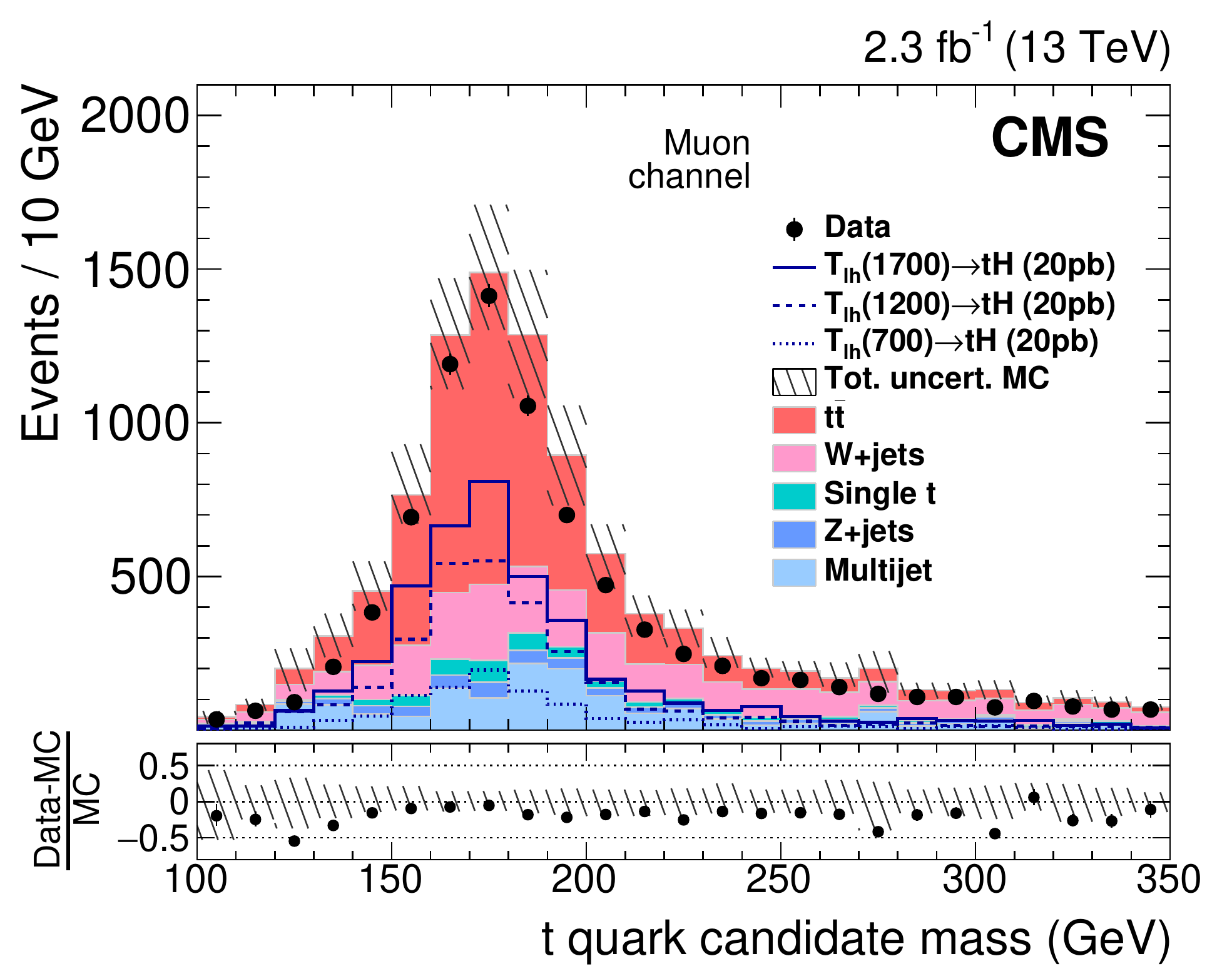}
\includegraphics[width=0.49\textwidth]{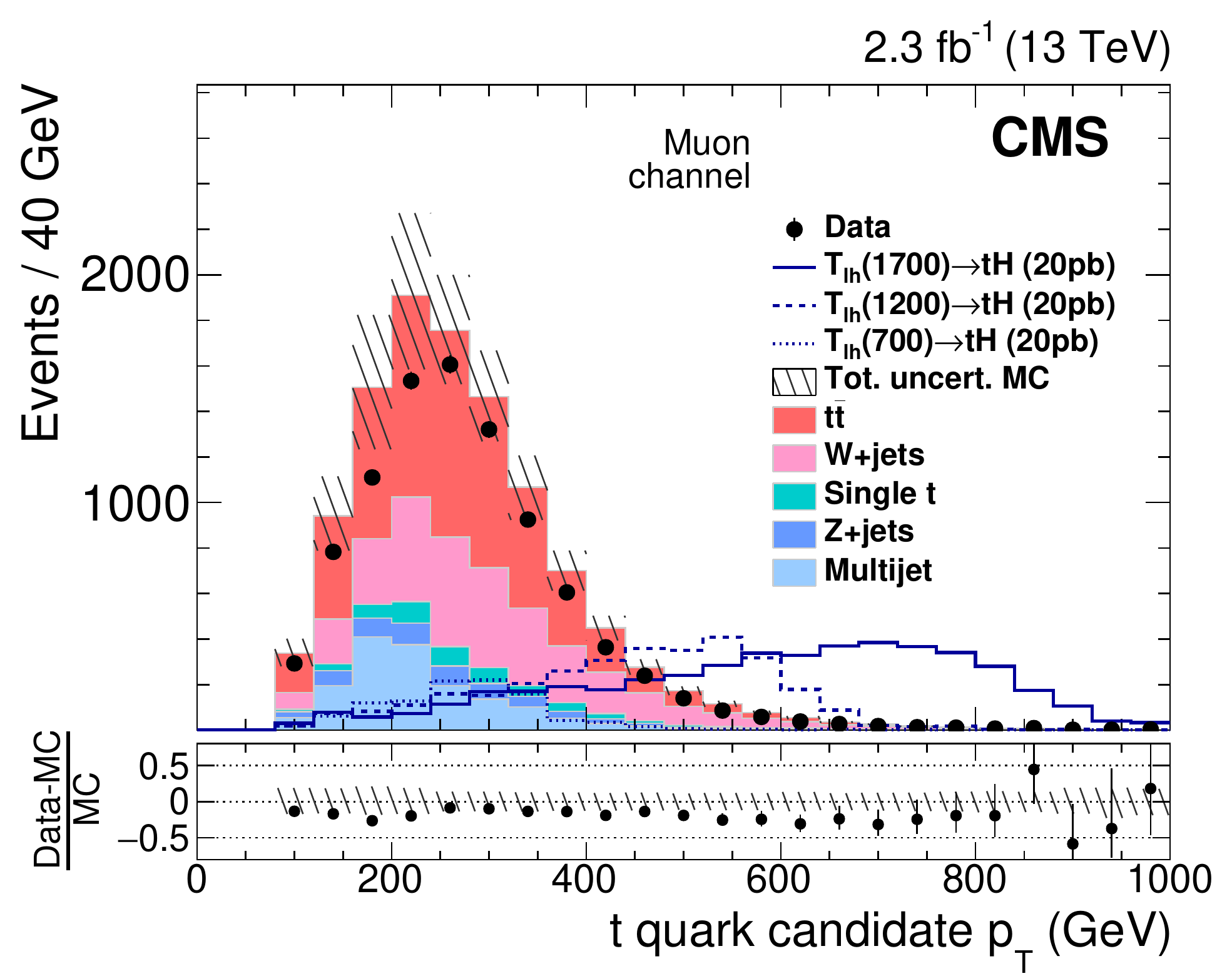}
\caption{
Mass (left) and \pt (right) distributions of the reconstructed top quark candidate in the muon channel after the baseline selection.
The different background contributions are shown using full histograms while the open histograms are signal yields and the data are shown as solid circles.
The hatched bands represent the statistical and systematic uncertainties of the simulated event samples.
The systematic uncertainties include those discussed in Section~\ref{sec_uncert}, except the forward jet uncertainty.
Signal cross sections are enhanced to 20\unit{pb}.
}
\label{fig_sel_t_H}
\end{figure*}

The selection criteria described above define the ``baseline selection''.
Distributions of some relevant variables after the baseline selection are shown in Figs.~\ref{fig_obj} and \ref{fig_sel_t_H}. The background contributions are estimated from simulated events. The hypothetical signal is scaled to a cross section of 20 pb as indicated in the legend of the figure. The simulated background events and data are found to be in agreement.

After the baseline selection, two event categories are defined. The signal region is used for signal extraction and is defined by requiring that both soft-drop subjets of the Higgs boson candidate are \PQb tagged and that there is at least one forward jet.
The ``control region'' for background estimation is defined by requiring the absence of forward jets and that exactly one of the soft-drop subjets of the Higgs boson candidates is \PQb tagged.
In addition, two validation regions with zero subjet b-tags, ``region A'' and ``region B'', are defined.
These validation regions are used to cross-check the background estimation method as described in Section~\ref{background}.
The event selection criteria of all regions are summarised in Table~\ref{tab_sig_sb_def}.

\begin{table*}[htb]
\centering
\topcaption{
Event selection criteria: required number of \PQb tagged subjets for the Higgs boson candidate, and number of forward jets.
}
\begin{tabular}{ l | c | c | c | c}
\hline
  & & & Validation & Validation \\
Region                                  & Signal region     & Control region  & region A & region B   \\
\hline
Subjet \PQb tags (H candidate)   & exactly 2         & exactly 1  & exactly 0 & exactly 0        \\
Forward jets                  & at least 1        & exactly 0    & exactly 0  & at least 1     \\
\hline
\end{tabular}
\label{tab_sig_sb_def}
\end{table*}

\begin{figure*}[htb]
\centering
\includegraphics[width=0.49\textwidth]{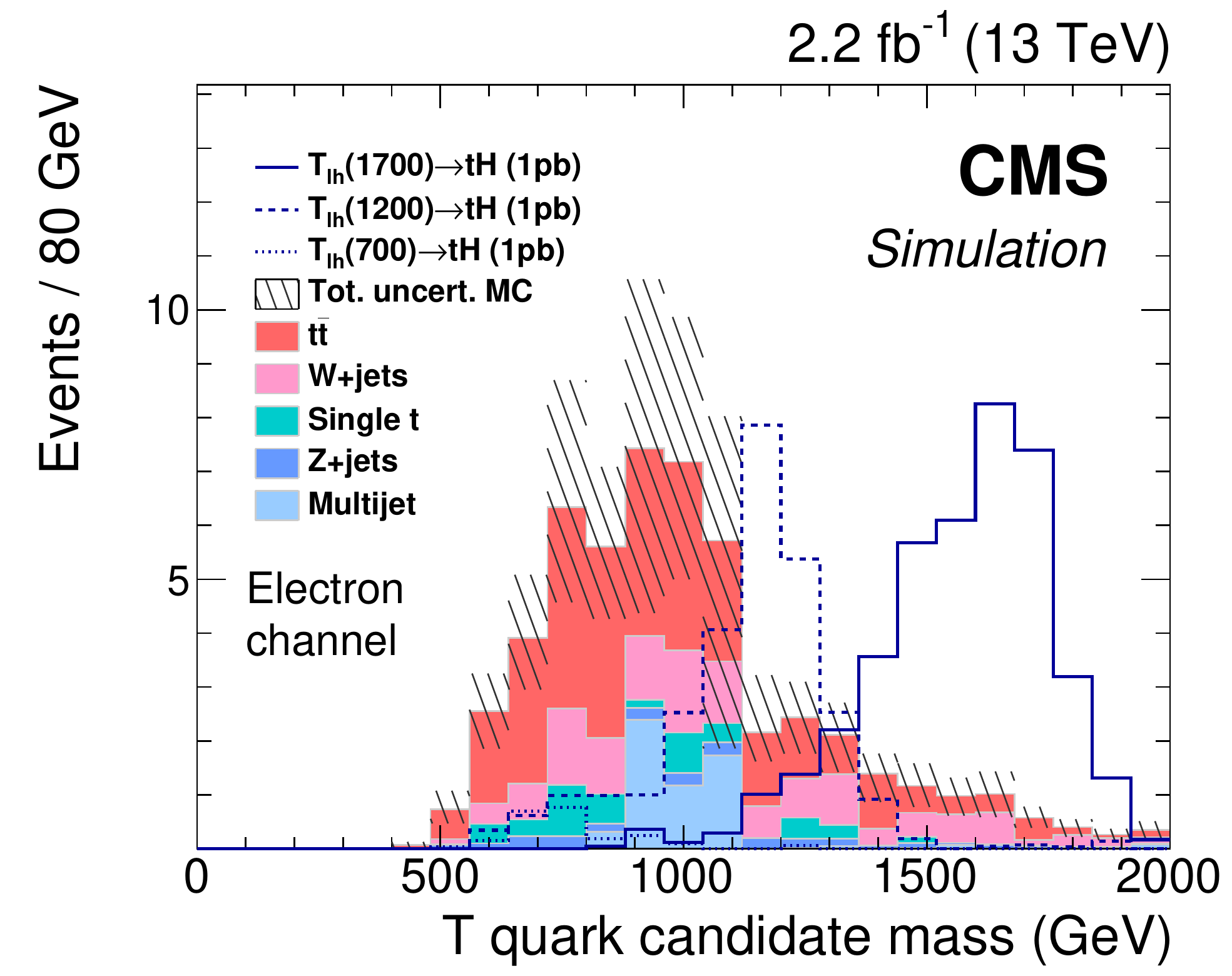}
\includegraphics[width=0.49\textwidth]{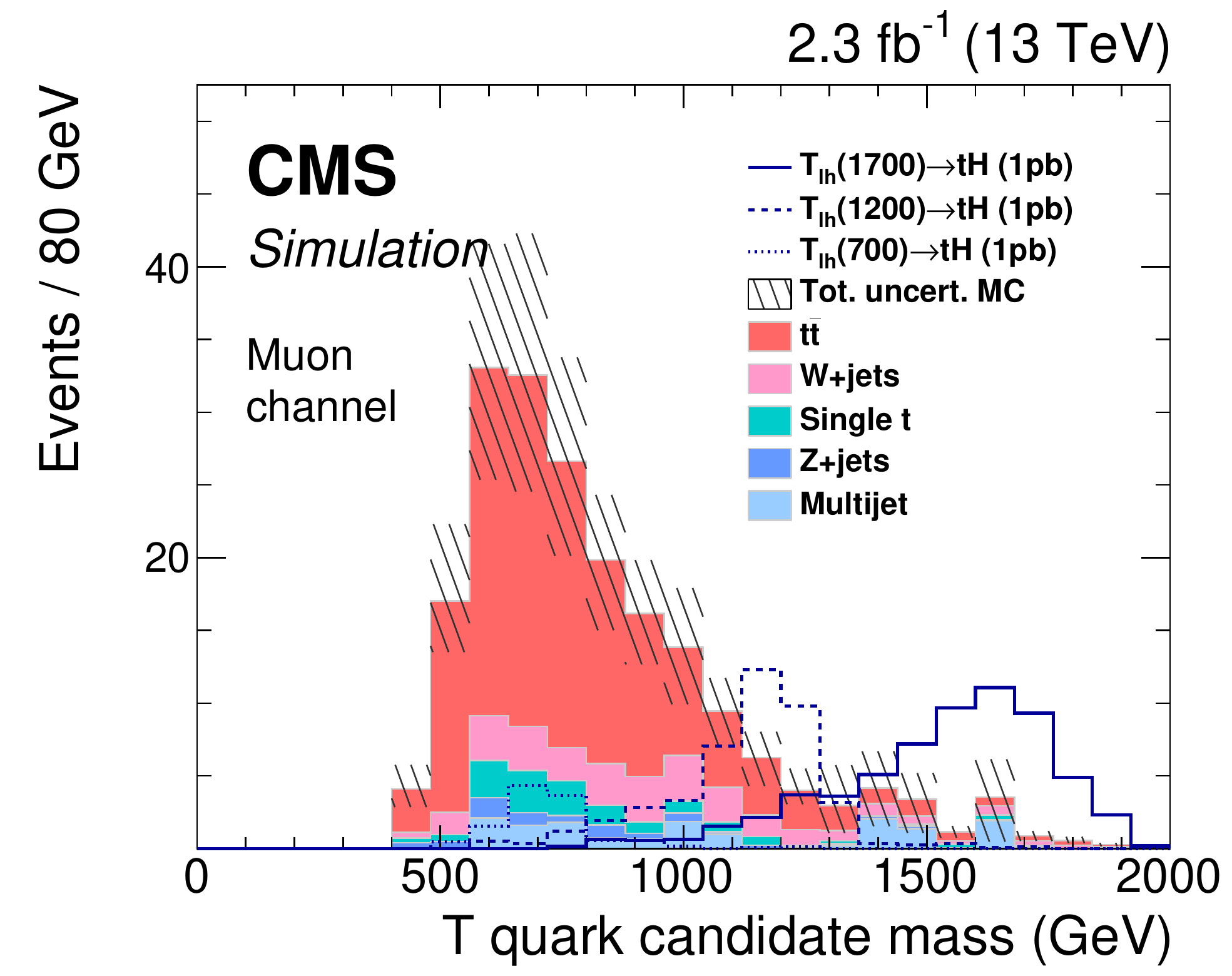}
\caption{
Vector-like \cPQT quark candidate mass in the signal region for the electron (left) and muon (right) channels.
The different background contributions are shown using full histograms while the open histograms are signal yields and the data are shown as solid circles.
The hatched bands represent the statistical and systematic uncertainties of the simulated event samples.
The systematic uncertainties include those discussed in Section~\ref{sec_uncert}.
}
\label{fig_sr_mass}
\end{figure*}

\begin{figure}[htb]
\centering
\includegraphics[width=0.49\textwidth]{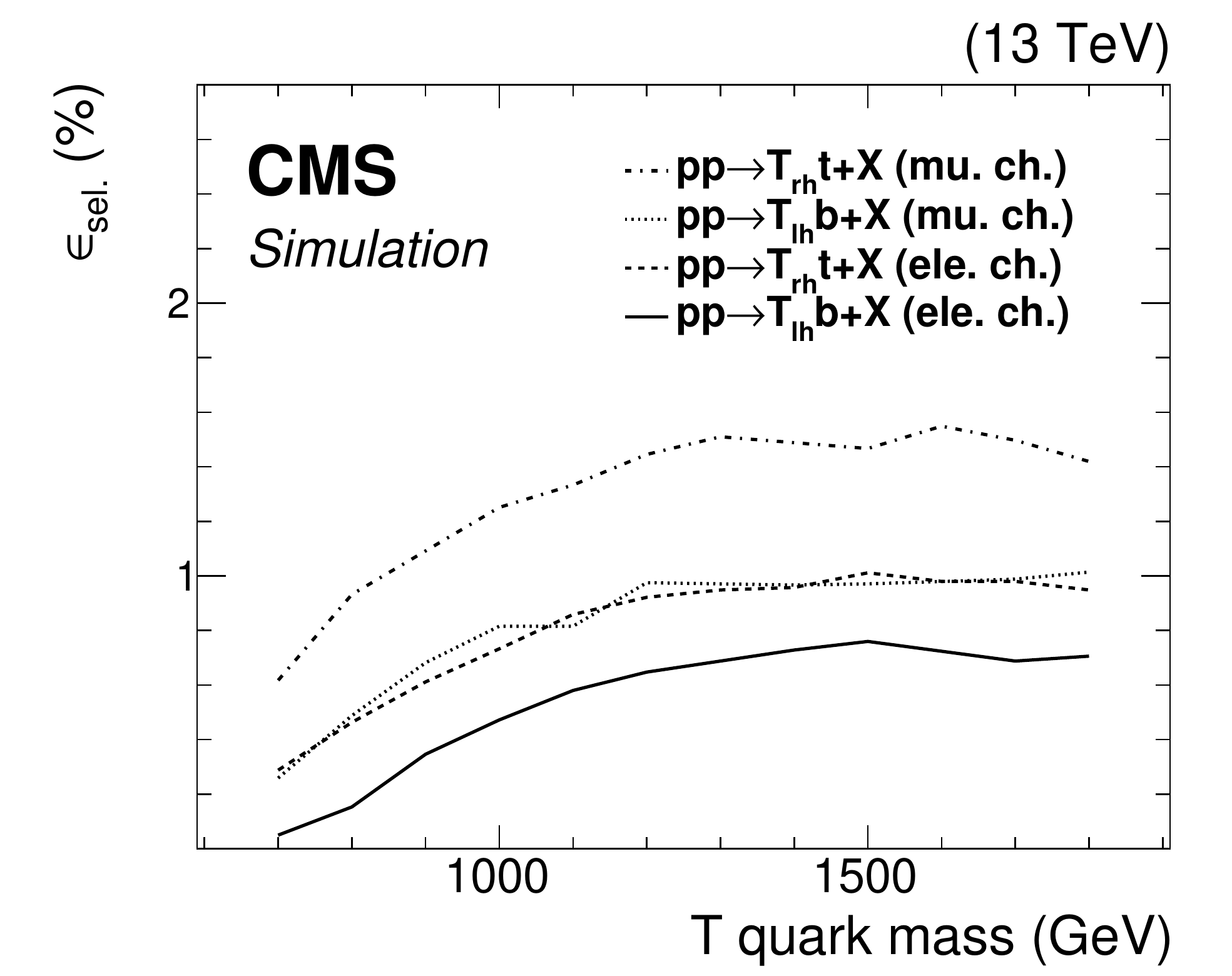}
\caption{
Selection efficiency $\epsilon_\text{sel}$ for the signal, i.e. the product of the branching fractions for the top quark decaying to final states including a lepton, and the Higgs boson decaying to bottom quarks, amounting to approximately 8\%, is included in the signal selection efficiency.
Left-handed (denoted by lh) and right-handed (denoted by rh) couplings of the \cPQT quark to SM particles in associated production with bottom and top quarks, respectively, are shown separately.
}
\label{fig_sel_eff}
\end{figure}

The \cPQT quark candidate is reconstructed from the sum of the Higgs boson and the top quark candidate four-momenta.
The \MT is used as the discriminating variable in the limit setting procedure.
Figure~\ref{fig_sr_mass} shows the simulated signal and background distributions of \MT in the signal region.
In the electron (muon) channel 35 (134) data events are selected, as summarised in Table~\ref{tab_evt_sel} along with the event yields and selection efficiencies for three of the signal samples.
The signal selection efficiency is depicted as a function of the generated \cPQT quark mass in Fig.~\ref{fig_sel_eff}.
The denominator of the efficiency includes all decay modes of the top quark and the Higgs boson, i.e. the product of the branching fractions for the top quark decaying to final states including a lepton, and the Higgs boson decaying to bottom quarks, amounting to approximately 8\%, is included in the signal selection efficiency.
The selection efficiency is notably larger for the right-handed signal samples, because of the harder \pt spectrum of leptons stemming from right-handed \cPQT quarks, and the presence of additional leptons from the associated top quark production.

\begin{table*}[htb]
\centering
\topcaption{
Number of selected events $N_\text{sel}$ and selection efficiency $\epsilon_\text{sel}$ for the signal region including both statistical (stat) and systematic (sys) uncertainties.
For the background, the post-fit value (as described in Sections~\ref{background} and \ref{sec_limits}) is quoted.
The left- (right-) handed \cPQT quark production in association with a bottom (top) quark is denoted by a subscript lh (rh) and following \PQb (\PQt).
All signal samples are normalized to a cross section of 1\unit{pb}, i.e. the product of the branching fractions for the top quark decaying to final states including a lepton, and the Higgs boson decaying to bottom quarks, amounting to approximately 8\%, is included in the signal selection efficiency.
}
\begin{tabular}{ l | c  c | c  c }
\hline
       & \multicolumn{2}{c|}{Electron channel}                                    & \multicolumn{2}{c}{Muon channel}                 \\
       & $N_\text{sel}\pm$stat$\pm$sys & $\epsilon_\text{sel}\,(\%)$ & $N_\text{sel}\pm$stat$\pm$sys & $\epsilon_\text{sel}\,(\%)$ \\
\hline
\cPQT$_\mathrm{lh}$(700)  \PQb & $  1.2 \pm  1.1 \pm  0.3$ & $0.05$ & $  6.0 \pm  2.4 \pm  1.2$ & $0.26$  \\
\cPQT$_\mathrm{lh}$(1200) \PQb & $ 14.4 \pm  0.9 \pm  2.6$ & $0.65$ & $ 22.8 \pm  1.1 \pm  3.9$ & $0.98$  \\
\cPQT$_\mathrm{lh}$(1700) \PQb & $ 15.3 \pm  0.9 \pm  2.7$ & $0.69$ & $ 22.9 \pm  1.1 \pm  3.9$ & $0.99$  \\
\hline
\cPQT$_\mathrm{rh}$(700)  \PQt & $  6.4 \pm  2.5 \pm  1.1$ & $0.29$ & $ 14.2 \pm  3.8 \pm  2.3$ & $0.61$  \\
\cPQT$_\mathrm{rh}$(1200) \PQt & $ 20.3 \pm  1.0 \pm  3.4$ & $0.91$ & $ 33.6 \pm  1.3 \pm  5.4$ & $1.45$  \\
\cPQT$_\mathrm{rh}$(1700) \PQt & $ 21.7 \pm  1.1 \pm  3.5$ & $0.98$ & $ 34.6 \pm  1.4 \pm  5.7$ & $1.49$  \\
\hline
                        & \multicolumn{2}{c|}{$N_\text{sel}\pm$stat$\pm$fit} & \multicolumn{2}{c}{$N_\text{sel}\pm$stat$\pm$fit}\\
\hline
Background (post-fit)   & \multicolumn{2}{c|}{$34.8 \pm  1.4 \pm  4.2$}     & \multicolumn{2}{c}{$133 \pm  3 \pm 16$}    \\
Data                    & \multicolumn{2}{c|}{35}            & \multicolumn{2}{c}{134}           \\
\hline
\end{tabular}
\label{tab_evt_sel}
\end{table*}

\section{Background Estimate}
\label{background}

The combined shape of the \MT distribution of all background processes is provided by the data in the control region.
It is used together with the simulated signal distribution in a fit of signal plus background distributions to the observed data.
The normalization of the background distribution is estimated in the fit.

\begin{figure*}[htbp]
\centering
\includegraphics[width=0.49\textwidth]{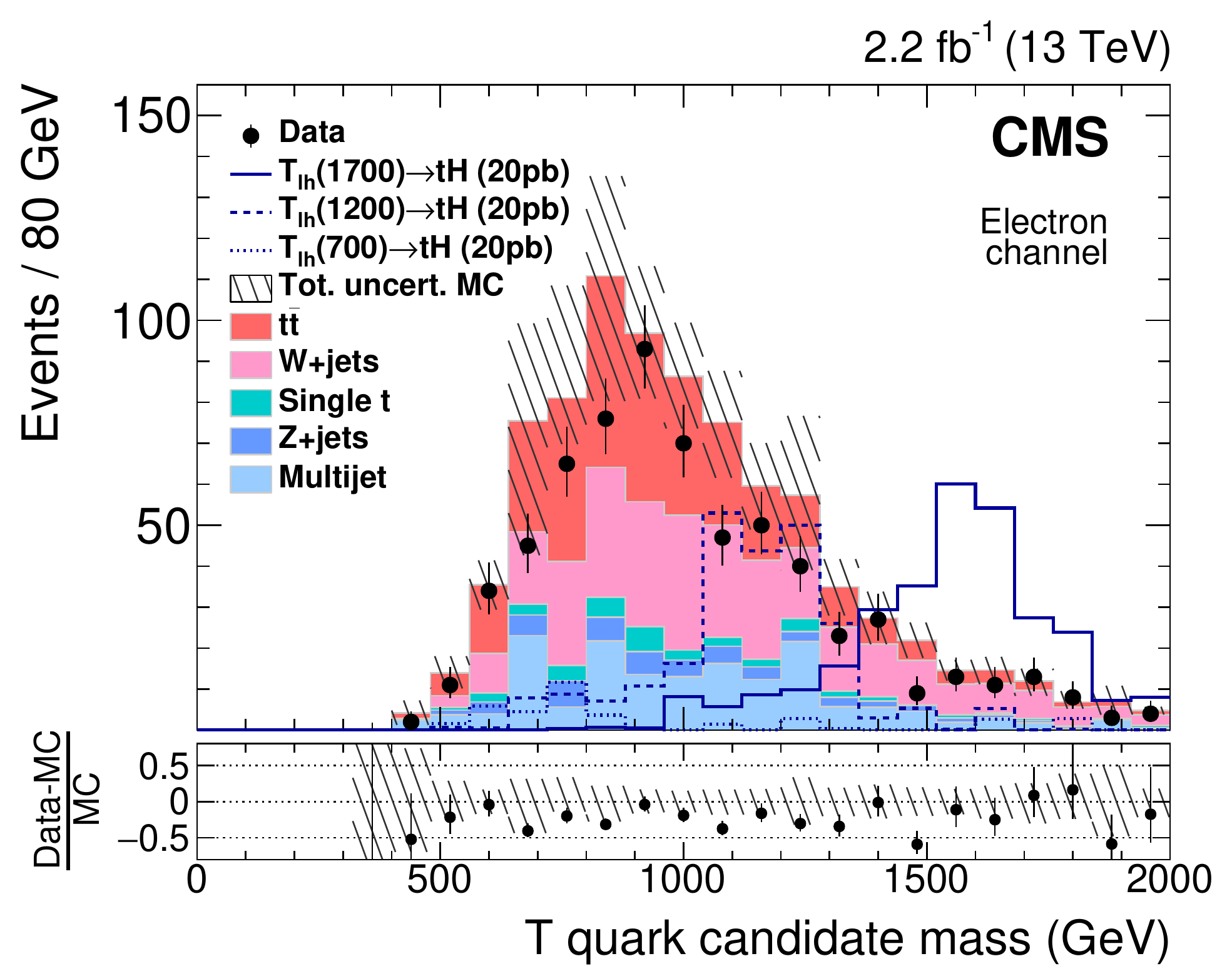}
\includegraphics[width=0.49\textwidth]{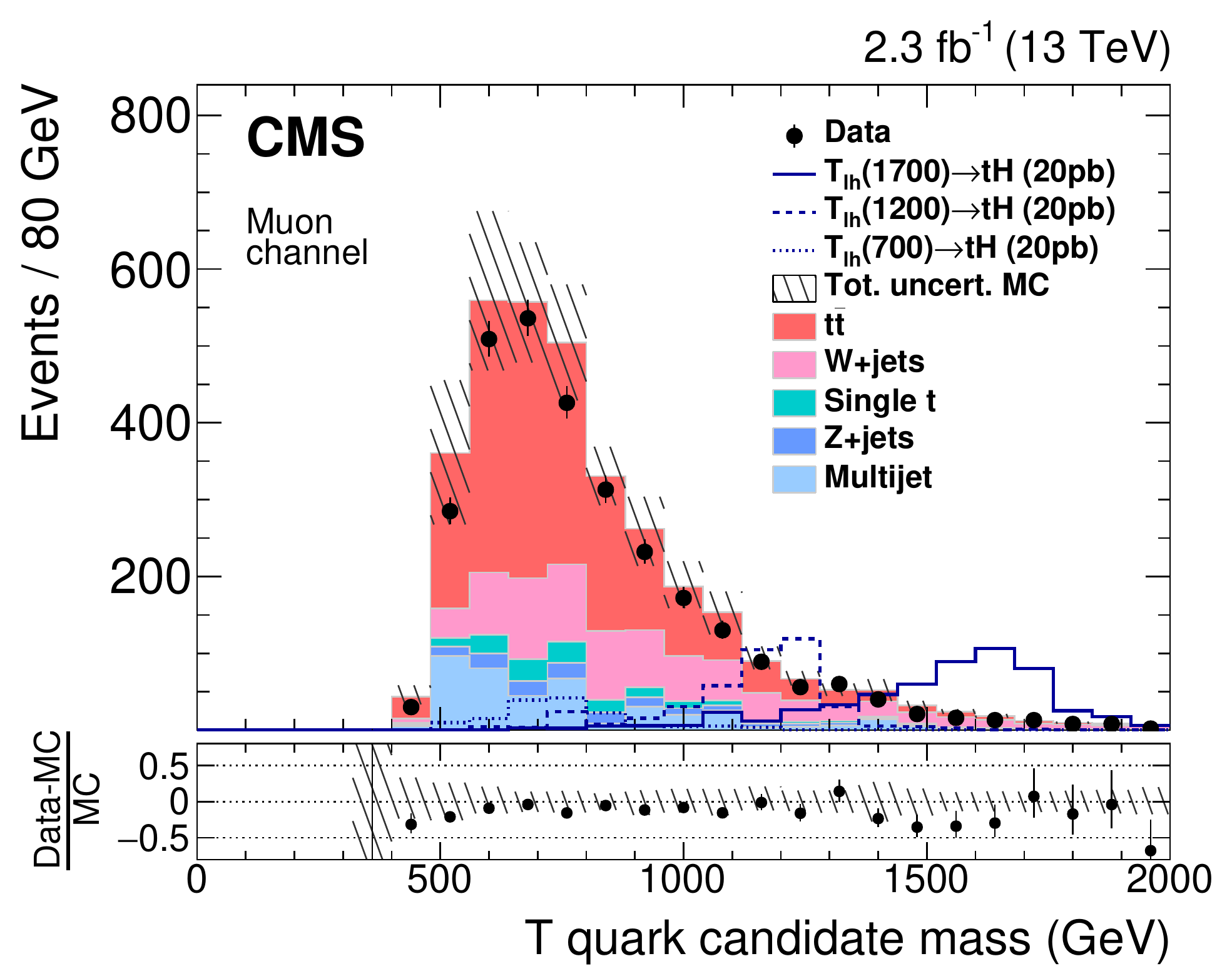}
\caption{
Vector-like \cPQT quark candidate mass in the control region for the electron (left) and muon (right) channels.
Signal samples are normalized to 20\unit{pb}, which is a factor of 20 larger than what is used in Figure \ref{fig_sr_mass}.
The shape of the data distribution provides the background estimate.
The different background contributions are shown using full histograms while the open histogram are signal yields and the data are shown as solid circles.
The hatched bands represent the statistical and systematic uncertainties of the simulated event samples.
The systematic uncertainties include those discussed in Section~\ref{sec_uncert}, except the forward jet uncertainty.
}
\label{fig_sb_mass}
\end{figure*}

Figure~\ref{fig_sb_mass} shows the reconstructed mass of the \cPQT quark candidates in the control region, where the signal cross section is increased by a factor of 20.
Data and simulation is observed to agree.
The control region features a signal-to-background ratio of approximately 5\% of that found in the signal region and can therefore be used to estimate the background with low signal contamination.

\begin{figure*}[htbp]
\centering
\includegraphics[width=0.49\textwidth]{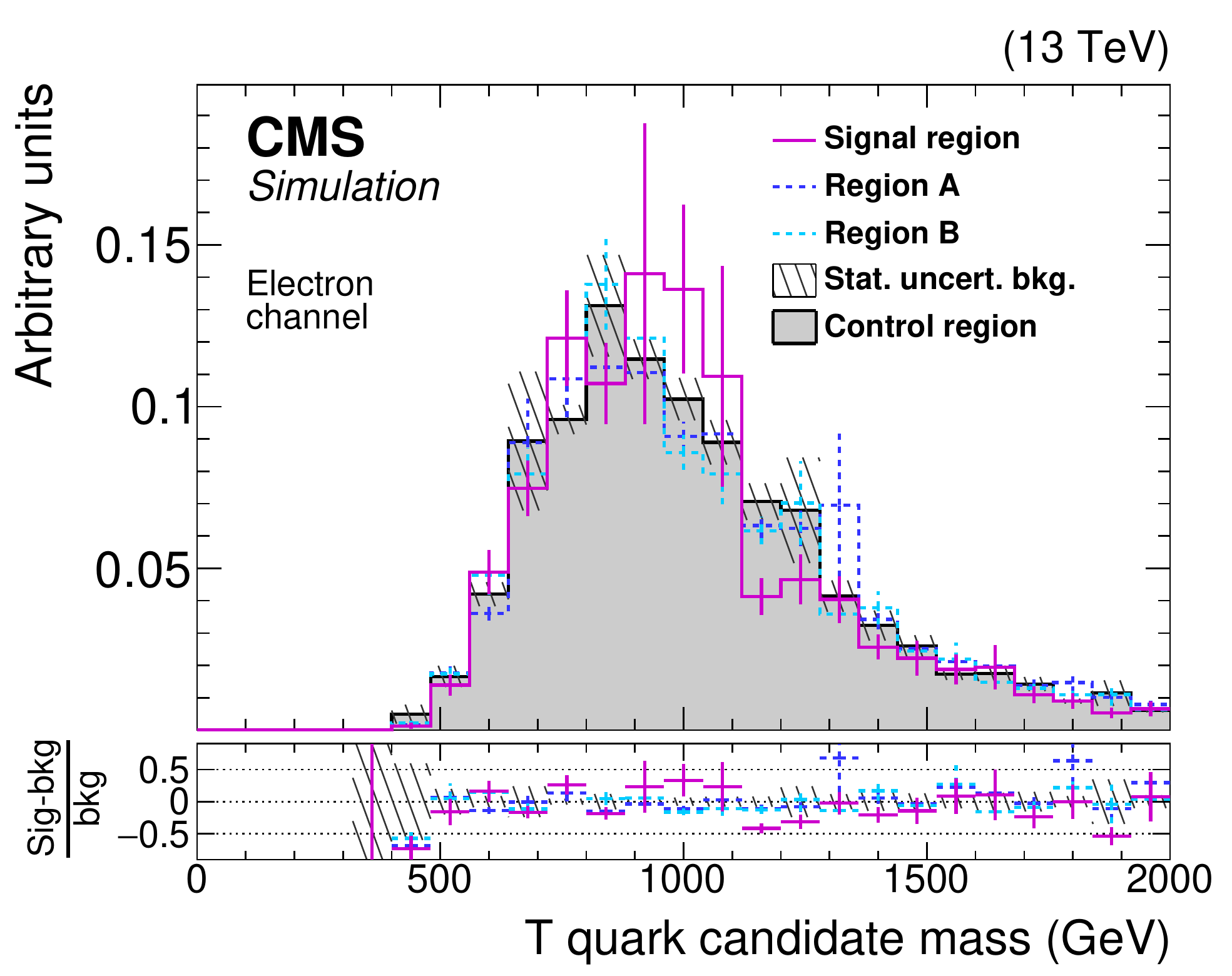}
\includegraphics[width=0.49\textwidth]{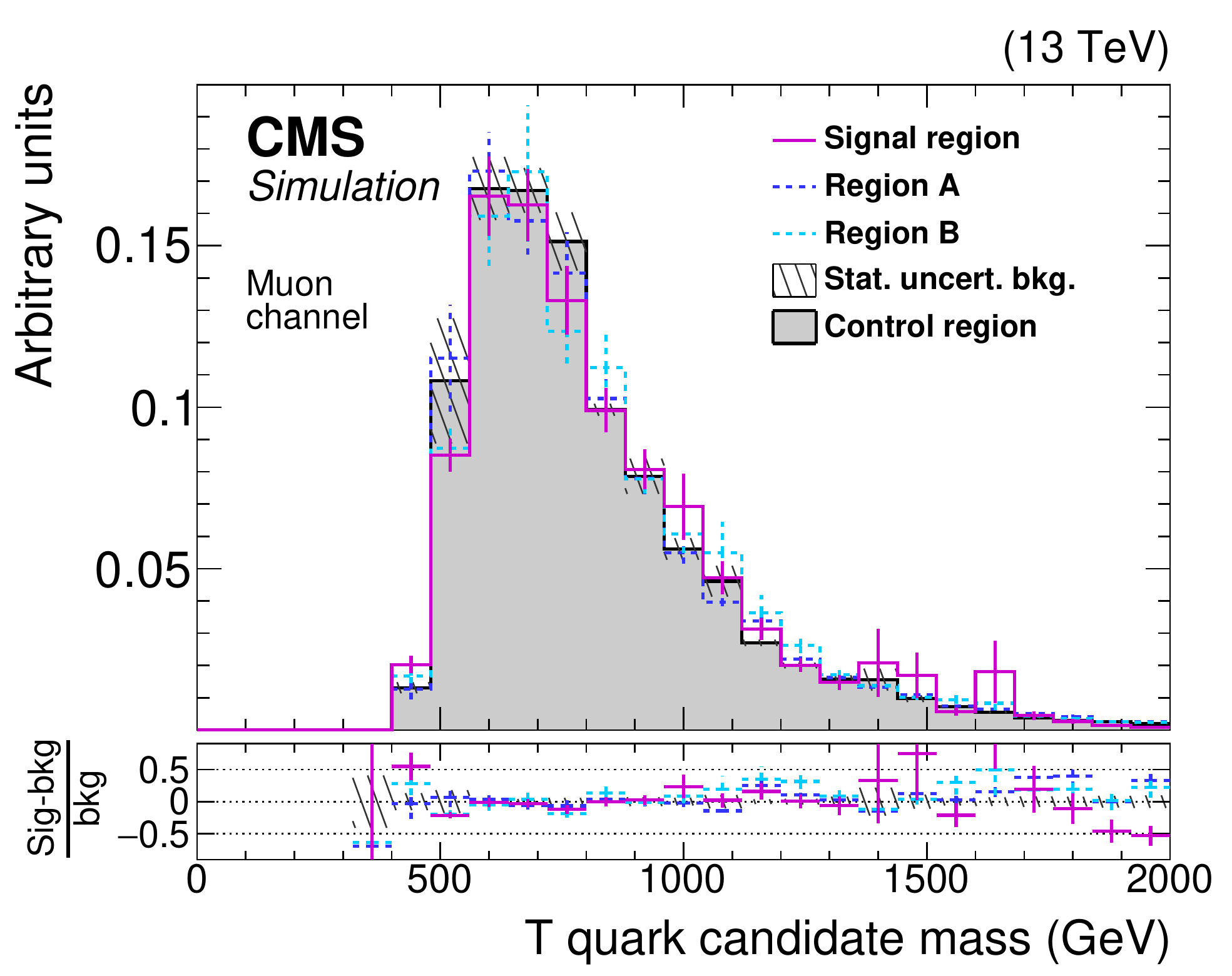}
\caption{
Shape comparison of the \cPQT quark candidate mass distributions in the signal (violet solid line) and control (shaded histogram) regions as well as the validation regions A (dark blue dashed line) and B (light blue dashed line) for the electron (left) and muon (right) channels.
The distributions show the sum of all simulated backgrounds, with the statistical uncertainties indicated as the error bars (signal region) or the hatched band (control region).
}
\label{fig_sig_sb_comp}
\end{figure*}

Both the signal and control regions contain 50--60\% top quark pair background and 20--30\% \PW+jets background.
The relative background composition is therefore similar in the signal and control regions.
Also the kinematic configuration of the top quark and Higgs boson candidates are similar.
These two features facilitate the derivation of the background shape from the control region in data without any further corrections.
This procedure is validated by a shape comparison of the \MT distribution between the signal and control regions in simulated events, as shown in Fig.~\ref{fig_sig_sb_comp}.
The compatibility of the distributions is evaluated with a $\chi^2$ test~\cite{2006physics...5123G}, including the statistical uncertainties of the simulation as weights in the test.
The p-values obtained in the electron and muon channels are 0.22 and 0.09, respectively.
Therefore, the \MT distributions are assumed to be compatible in the signal and control regions. In addition Fig.~\ref{fig_sig_sb_comp} shows further cross checks using zero subjet \PQb tags on the Higgs boson candidate, thereby enriching the contribution of the \PW+jets and QCD backgrounds. Also these regions are in good agreement.

The aforementioned shape comparison is repeated for systematic uncertainties that can change the shape of the \MT distribution in either the signal or the control region.
These are the jet energy scale and resolution uncertainties, as well as uncertainties in the \PQb tag status of a Higgs boson candidate subjet.
Background cross sections are varied by twice their uncertainty, except for the multijet background, which is varied by half the estimated value.
Each variation in a systematic uncertainty is applied consistently in both regions.

\begin{figure*}[htbp]
\centering
\includegraphics[width=0.49\textwidth]{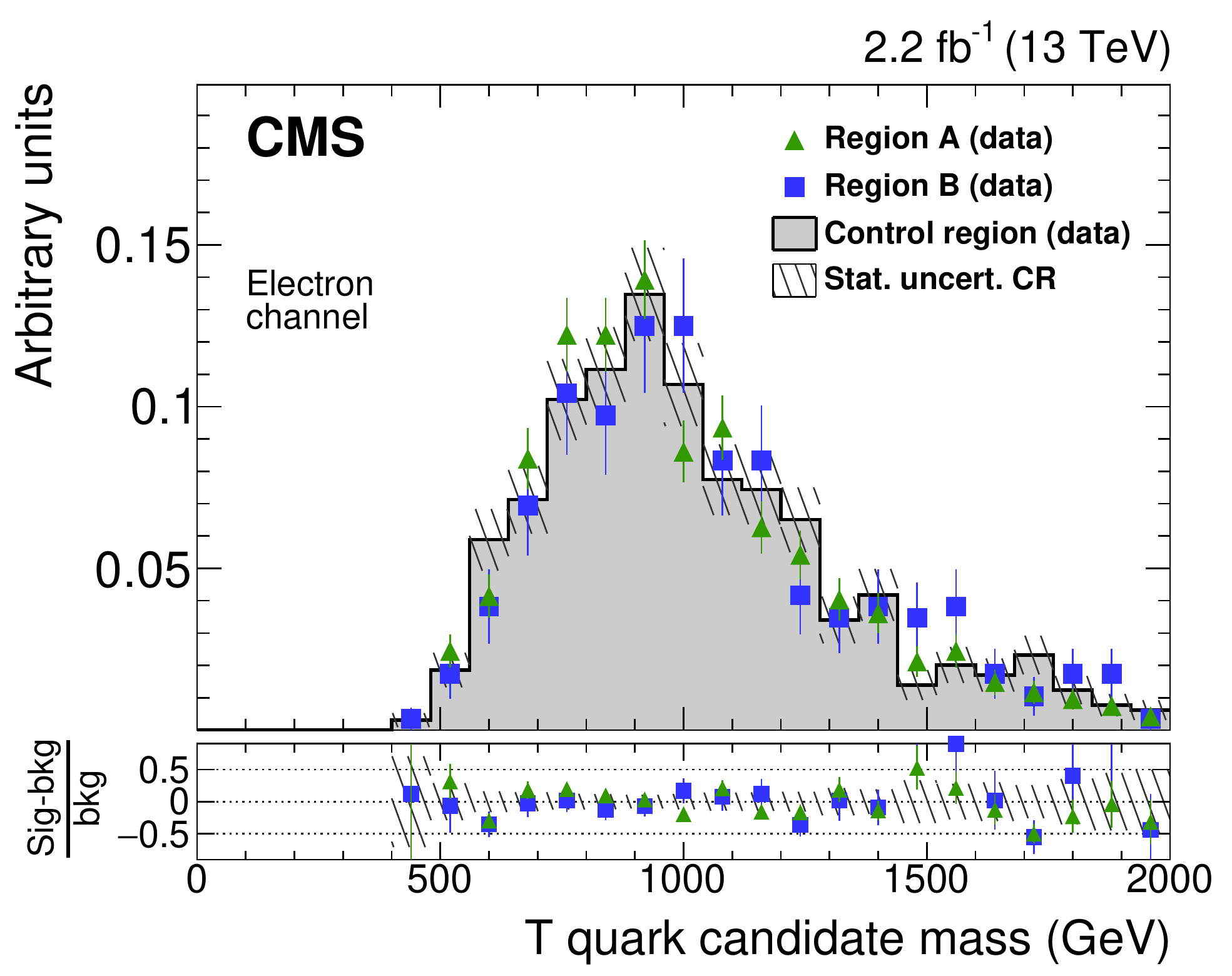}
\includegraphics[width=0.49\textwidth]{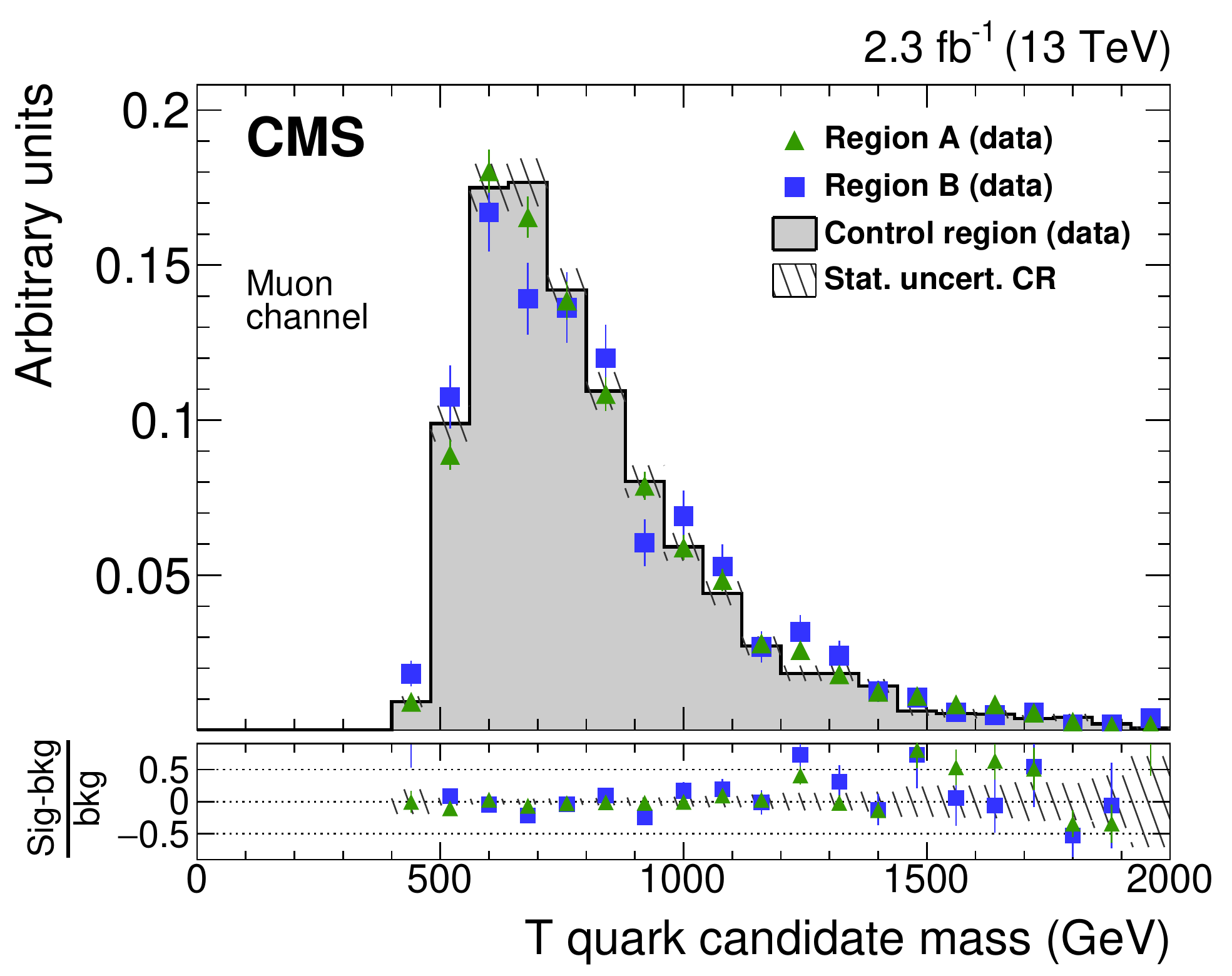}
\caption{
Shape comparison of the \cPQT quark candidate mass distributions in the control region (shaded histogram) regions and the validation regions A (green) and B (blue) for the electron (left) and muon (right) channels in data.
}
\label{fig_data_validationAB}
\end{figure*}

The compatibility between the validation regions and the control region is also checked in data, as shown in Fig.~\ref{fig_data_validationAB}.
Agreement between the corresponding regions is observed in all cases.

\begin{figure*}[htbp]
\centering
\includegraphics[width=0.49\textwidth]{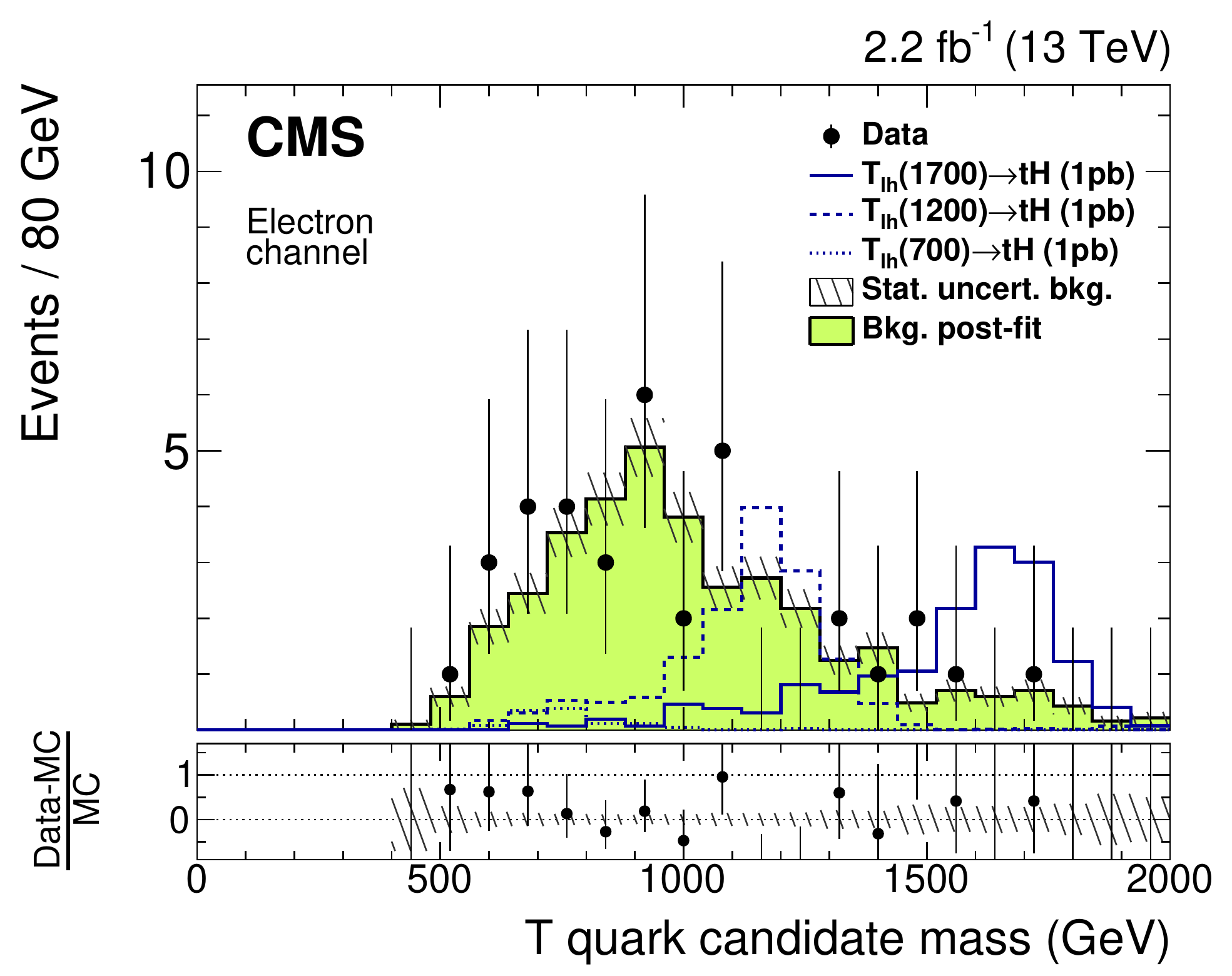}
\includegraphics[width=0.49\textwidth]{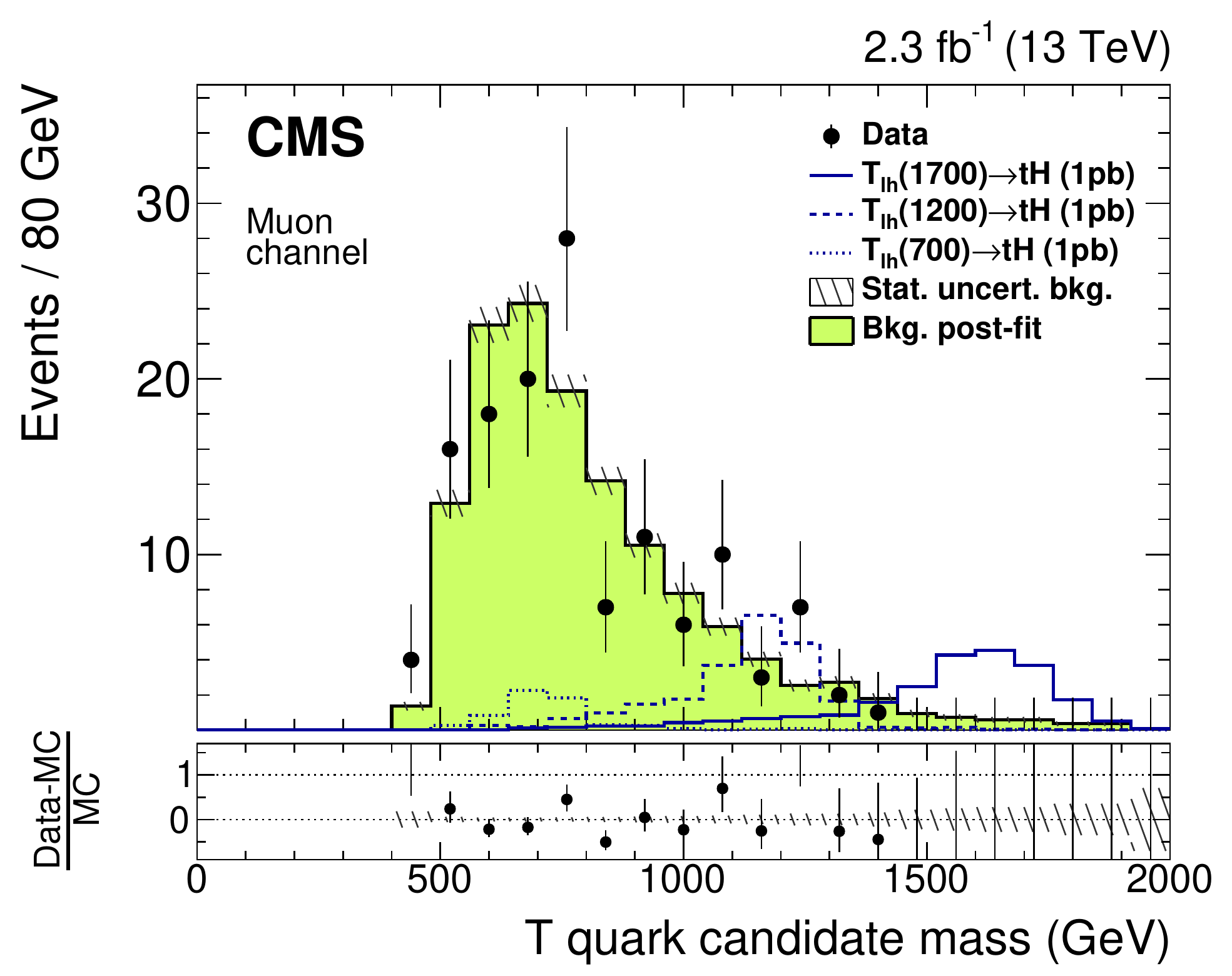}
\caption{
Final background, data, and expected signal distributions in \MT in the signal region for the electron (left) and muon (right) channels.
The hatched uncertainty band shows the statistical uncertainty in the background prediction, which is used as the shape uncertainty in the fit, as detailed in Section~\ref{sec_uncert}.
The normalization of the background estimate is taken from the fit, its uncertainty is 12\% (not included in the hatched uncertainty band).
}
\label{fig_money_plot}
\end{figure*}

In the control region, 632 (2949) events are selected in the electron (muon) channel. These relatively large numbers of events ensure that the statistical uncertainty is negligible compared to that in the signal region.
In Fig.~\ref{fig_money_plot} the background estimate is shown with the distribution of \MT in data.

\section{Systematic Uncertainties}
\label{sec_uncert}

Sources of systematic uncertainty may influence the rate and shape of the signal predictions as well as the shape of the background distribution.
The background shape uncertainty is taken as the uncertainty in each bin of the distribution of its estimate.
Note that there is no rate uncertainty associated with the background prediction described in Section \ref{background}, since its normalization is not used to obtain the final results.
In the above figures, several rate and shape uncertainties are considered, for the simulations of both the signal and the background.
The one with the largest effect on the final result originates from the uncertainty in the forward jet selection efficiency.
The next largest contributions arise from the uncertainties in the \PQb tag efficiency and jet energy corrections.
The impacts of the systematic uncertainties on the event rates are listed in Table~\ref{tab_sys_uncerts}.

\begin{table*}[hbtp]
\centering
\topcaption{
Impacts of the largest systematic uncertainties in the signal event yields.
The signal samples for $\cPQT_\mathrm{lh}\PQb$ production are shown.
The uncertainties in the forward jet, and lepton isolation and trigger are rate uncertainties, all other uncertainties are evaluated bin-by-bin.
All values are reported as percentage of the signal event yield.
}
\begin{tabular}{l|lll|lll}
\hline
                   & \multicolumn{3}{c|}{Electron channel}                         & \multicolumn{3}{c}{Muon channel}                             \\
                   & ${\cPQT_\mathrm{lh}(700) }$      & ${\cPQT_\mathrm{lh}(1200)}$   & ${\cPQT_\mathrm{lh}(1700)}$   & ${\cPQT_\mathrm{lh}(700) }$   & ${\cPQT_\mathrm{lh}(1200)}$   & ${\cPQT_\mathrm{lh}(1700)}$   \\
\hline
\PQb tagging, heavy flavour     & 7.8   & 7.6   & 8.7   & 6.0   & 7.5   & 8.5   \\
\PQb tagging, light flavour     & 0.7   & 0.7   & 0.5   & 1.2   & 0.6   & 0.7   \\
Forward jet                     & 15    & 15    & 15    & 15    & 15    & 15    \\
Jet energy resolution           & 7.0   & 0.6   & 0.2   & 0.2   & 2.3   & 0.9   \\
Jet energy scale                & 9.0   & 4.2   & 4.9   & 3.8   & 3.8   & 4.4   \\
Lepton isolation and trigger    & 5.0   & 5.0   & 5.0   & 5.0   & 5.0   & 5.0   \\
Soft-drop mass                  & 3.1   & 1.1   & 0.3   & 0.5   & 0.3   & 1.3   \\
PDF                             & 4.8   & 2.7   & 4.2   & 4.8   & 2.8   & 4.1   \\
Luminosity                      & 2.7   & 2.7   & 2.7   & 2.7   & 2.7   & 2.7   \\
Pileup                          & 1.4   & 0.6   & 0.1   & 1.3   & 0.7   & 1.1   \\
\hline
\end{tabular}
\label{tab_sys_uncerts}
\end{table*}

Scale factors for the \PQb tagging efficiency  are applied to simulated events to match the \PQb tagging performance observed in data~\cite{Chatrchyan:2012jua}.
The scale factors have a systematic uncertainty of 2--5\% for jets originating from \PQb hadrons, 4--10\% for \PQc quark jets and 7--10\% for light-flavour jets, all depending on the \pt of the jet.
Those uncertainties are propagated to the final result, where the uncertainties for heavy-flavour (\PQb and \PQc) jets and light-flavour (\PQu, \PQd, \PQs, \Pg) jets are treated as correlated within their group, but the uncertainties for heavy-flavour jets are assumed to be uncorrelated with those for light-flavour jets.

Jet energy scale and resolution corrections depend on the jet \pt and $\eta$.
The associated uncertainties are typically a few percent.
The resulting uncertainty in the signal yield is derived by applying the ${\pm}1\sigma$ variations simultaneously to AK4 and AK8 jets and also propagating the variation of jet momenta into the calculation of \ETmiss at the same time.
The ${\pm}1\sigma$ variations for the resolution smearing in the soft-drop mass are evaluated separately.
Additionally, as the reconstruction efficiency of forward jets has been observed to be larger in the simulation compared to the data, a rate uncertainty of $\pm$15\% is assigned to the signal samples.
This uncertainty is determined by evaluating the event selection efficiency using forward jets in two control regions requiring an event to be selected by the baseline selection and additionally having either zero subjet \PQb tags or exactly one, in association with the \PH boson candidate.
The central region is well modelled by the simulation.

To estimate the uncertainty in the pileup simulation, a variation of $\pm$5\% in the inelastic cross section value~\cite{atlas_pp_xsec}, controlling the average pileup multiplicity, is used.
The uncertainty in the luminosity measurement is $\pm$2.7\%~\cite{CMS-PAS-LUM-15-001}.
Systematic identification and trigger uncertainties for electrons and muons are taken into account for the signal processes.
The combined trigger and lepton isolation ($\Delta R (\ell, j)$ or $\pt^\text{rel}(\ell, j)$) selection efficiency has a rate uncertainty of $\pm$5\%.
For the PDF uncertainty the complete set of NNPDF~3.0 PDF eigenvectors are evaluated, following the PDF4LHC prescription~\cite{0954-3899-43-2-023001}.

\section{Results}
\label{sec_limits}

No significant deviation is observed from the shape predicted by the SM (see Fig.~\ref{fig_money_plot}).
The p-values of the compatibility tests between the predicted and observed distributions are 0.97 and 0.51 in the electron and muon channels, respectively.

Exclusion limits are set on the product of the production cross section and the branching fraction for single production of a vector-like \cPQT quark decaying to a top quark and a Higgs boson.
The 95\% confidence level (CL) exclusion limits are derived with a Bayesian statistical method~\cite{theta-stat, bayesbook}, where background and signal templates in the \MT distribution are used to make a combined fit to the data in the electron and muon channels.
Systematic uncertainties are included as nuisance parameters.
For rate-only uncertainties a log-normal prior is assigned.
A flat prior is used for the signal strength.
Shape uncertainties in the signal templates are taken into account using template morphing with cubic-linear interpolation, where the cubic interpolation is used up to the one sigma deviation and the linear interpolation beyond that.
For the background normalization a Gaussian prior with 100\% width is used.
The statistical uncertainty in the background estimate is included with the ``Barlow-Beeston light'' method~\cite{Barlow1993}, which uses a Gaussian approximation of the uncertainty in each bin.
A bias-test is performed by injecting a signal into the fitted data. The biases are observed to be negligible.

The obtained exclusion limits are compared with predictions from two benchmark models.
For $\cPQT_\mathrm{lh}\PQb$ production, branching fractions of 50/25/25\% for the \cPQT quark decay to \bW/\tZ/\tH are considered. These branching fractions correspond to the predictions for a vector-like isospin singlet.
A scenario with neutral currents only and equal couplings to \tZ and \tH is used for $\cPQT_\mathrm{rh}\PQt$ production (0/50/50\%), corresponding to the prediction for an isospin doublet.
Signal cross sections are taken from NLO calculations~\cite{Matsedonskyi2014, PhysRevD.70.094012} and multiplied with a factor of 0.25 and 0.5 in order to accommodate the branching fraction $\mathcal{B}(\PQt\PH)=\mathcal{B}(\PQb\PW)/2$ and $\mathcal{B}(\PQt\PH)=\mathcal{B}(\PQt\Z)$ for $\cPQT_\mathrm{lh}\PQb$ and $\cPQT_\mathrm{rh}\PQt$ production, respectively.
Single vector-like quark production is parametrised with a coupling constant to the electroweak sector.
For the coupling of a left- (right-) handed \cPQT quark to a quark and boson
pair, qV, the coupling strength, as defined in
Ref.~\cite{Matsedonskyi2014}, of $c^{\PQb\PW\,(\PQt\Z)}_\mathrm{L\,(R)}=0.5$ is assumed in production, where $c$ is a factor multiplying the weak coupling constant $ g_\mathrm{w}$.
For a coupling parameter of 0.5, it has been verified that the experimental resolution is much larger than the width of the \cPQT quark in the simplified model.

In the simplest Simplified Model~\cite{Matsedonskyi2014}, only the left- (right-) handed couplings are allowed for the singlet (doublet) scenarios, i.e.  $c^{\PQb\PW\,(\PQt\Z)}_\mathrm{R\,(L)}=0$, simultaneously for production and decay of the \cPQT quark.
Therefore, only fully left- (right-) handed polarisations are considered for the exclusion limits.

\begin{figure*}[htbp]
\centering
\includegraphics[width=0.49\textwidth]{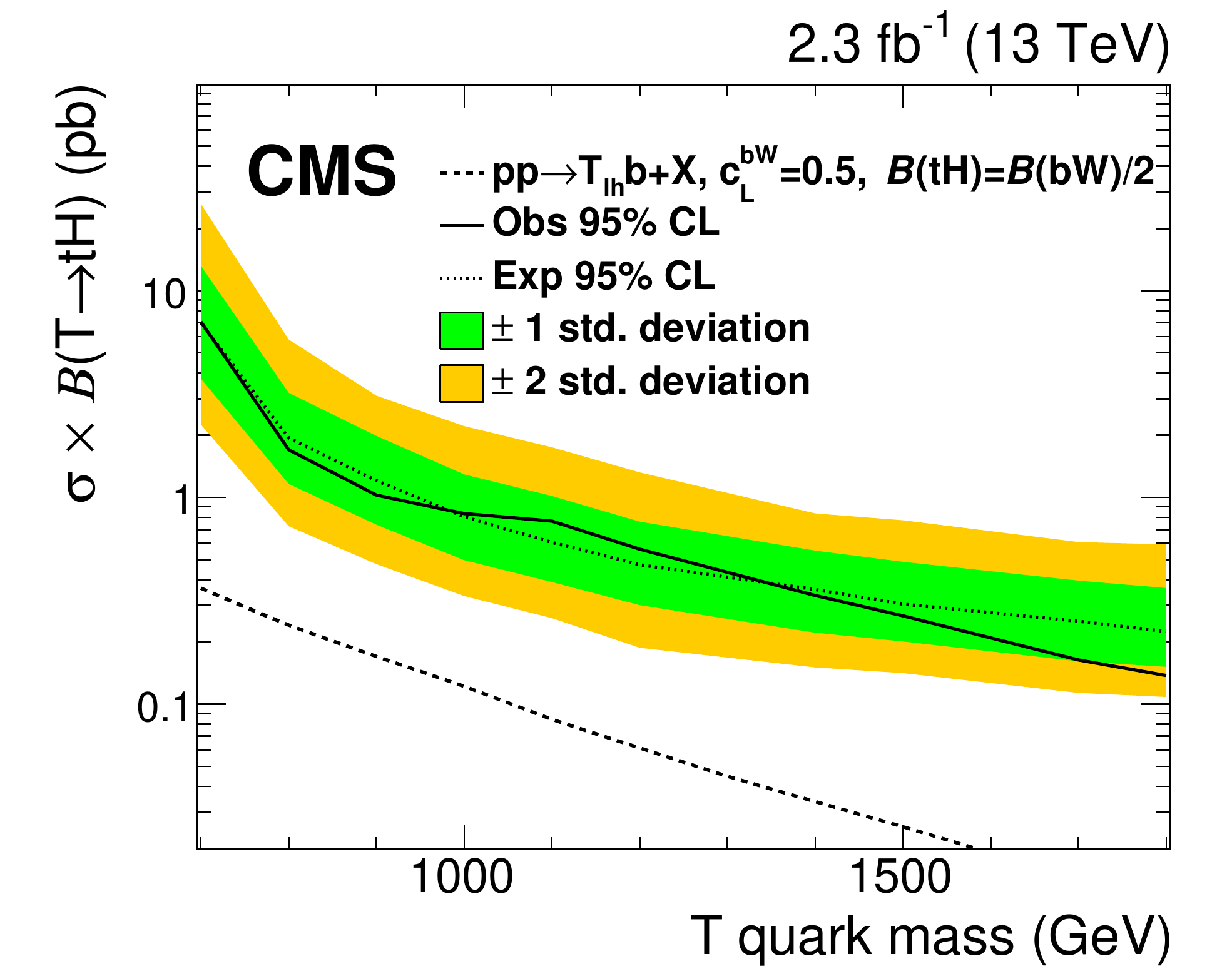}
\includegraphics[width=0.49\textwidth]{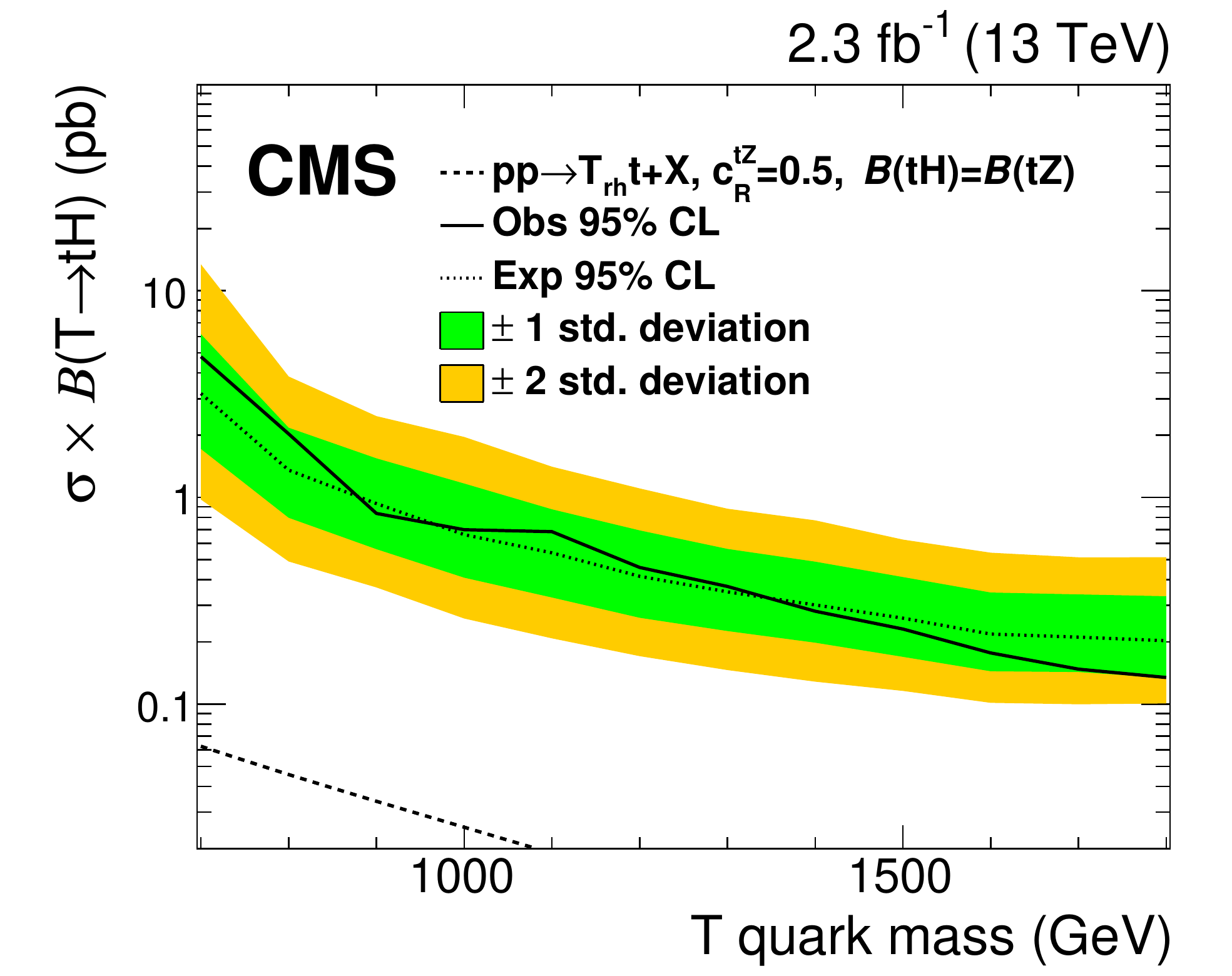}
\caption{
Exclusion limits on the product of the cross section and the branching fraction of single \cPQT quark production  and $\cPQT\to \PQt\PH$ decay.
A simultaneous fit is made to the electron and muon channels.
Left- (right-) handed \cPQT quark production in association with a bottom (top) quark is shown in the left- (right-) diagram.
}
\label{fig_limits_comb}
\end{figure*}

Figure \ref{fig_limits_comb} shows the 95\% CL upper limits on the product of the cross section and the branching fraction, along with the predictions of the simplest Simplified Model with coupling to third generation SM quarks only.
It can be seen that the excluded cross sections are an order of magnitude higher than the predictions, and the current data do not place constraints on this particular model.
This is the first search for singly produced VLQ by the CMS Collaboration.
In the future, results in this channel will become more sensitive by combining results with other final states, and it is anticipated that such Simplified Model cross sections will be probed with the large expected LHC Run 2 dataset.

\section{Summary}

A search for a singly produced vector-like \cPQT quark decaying to a top quark and a Higgs boson has been presented, where the top quark decay includes an electron or a muon and the Higgs boson decays into a pair of \PQb quarks.
For every event, the four-momentum of the vector-like \cPQT quark candidate is reconstructed and its mass is evaluated.
No excess over the estimated backgrounds is observed.
Upper limits are placed on the product of the cross section and the branching fraction for vector-like \cPQT quarks to a top quark and a Higgs boson in the mass range of 700 to 1800\GeV, at 95\% confidence level.
For a \cPQT quark with a mass of 1000\GeV with left- (right-) handed coupling to standard model particles, we exclude a value of the product of the production cross section and the branching fraction greater than 0.8 (0.7)\unit{pb}.
This is the first analysis setting exclusion limits on the cross section of singly produced vector-like \cPQT quarks at a centre-of-mass energy of 13\TeV.

\begin{acknowledgments}
We congratulate our colleagues in the CERN accelerator departments for the excellent performance of the LHC and thank the technical and administrative staffs at CERN and at other CMS institutes for their contributions to the success of the CMS effort. In addition, we gratefully acknowledge the computing centres and personnel of the Worldwide LHC Computing Grid for delivering so effectively the computing infrastructure essential to our analyses. Finally, we acknowledge the enduring support for the construction and operation of the LHC and the CMS detector provided by the following funding agencies: BMWFW and FWF (Austria); FNRS and FWO (Belgium); CNPq, CAPES, FAPERJ, and FAPESP (Brazil); MES (Bulgaria); CERN; CAS, MoST, and NSFC (China); COLCIENCIAS (Colombia); MSES and CSF (Croatia); RPF (Cyprus); SENESCYT (Ecuador); MoER, ERC IUT, and ERDF (Estonia); Academy of Finland, MEC, and HIP (Finland); CEA and CNRS/IN2P3 (France); BMBF, DFG, and HGF (Germany); GSRT (Greece); OTKA and NIH (Hungary); DAE and DST (India); IPM (Iran); SFI (Ireland); INFN (Italy); MSIP and NRF (Republic of Korea); LAS (Lithuania); MOE and UM (Malaysia); BUAP, CINVESTAV, CONACYT, LNS, SEP, and UASLP-FAI (Mexico); MBIE (New Zealand); PAEC (Pakistan); MSHE and NSC (Poland); FCT (Portugal); JINR (Dubna); MON, RosAtom, RAS, and RFBR (Russia); MESTD (Serbia); SEIDI and CPAN (Spain); Swiss Funding Agencies (Switzerland); MST (Taipei); ThEPCenter, IPST, STAR, and NSTDA (Thailand); TUBITAK and TAEK (Turkey); NASU and SFFR (Ukraine); STFC (United Kingdom); DOE and NSF (USA).

\hyphenation{Rachada-pisek} Individuals have received support from the Marie-Curie programme and the European Research Council and EPLANET (European Union); the Leventis Foundation; the A. P. Sloan Foundation; the Alexander von Humboldt Foundation; the Belgian Federal Science Policy Office; the Fonds pour la Formation \`a la Recherche dans l'Industrie et dans l'Agriculture (FRIA-Belgium); the Agentschap voor Innovatie door Wetenschap en Technologie (IWT-Belgium); the Ministry of Education, Youth and Sports (MEYS) of the Czech Republic; the Council of Science and Industrial Research, India; the HOMING PLUS programme of the Foundation for Polish Science, cofinanced from European Union, Regional Development Fund, the Mobility Plus programme of the Ministry of Science and Higher Education, the National Science Center (Poland), contracts Harmonia 2014/14/M/ST2/00428, Opus 2014/13/B/ST2/02543, 2014/15/B/ST2/03998, and 2015/19/B/ST2/02861, Sonata-bis 2012/07/E/ST2/01406; the Thalis and Aristeia programmes cofinanced by EU-ESF and the Greek NSRF; the National Priorities Research Program by Qatar National Research Fund; the Programa Clar\'in-COFUND del Principado de Asturias; the Rachadapisek Sompot Fund for Postdoctoral Fellowship, Chulalongkorn University and the Chulalongkorn Academic into Its 2nd Century Project Advancement Project (Thailand); and the Welch Foundation, contract C-1845.
\end{acknowledgments}
\vspace*{2em}
\bibliography{auto_generated}

\cleardoublepage \appendix\section{The CMS Collaboration \label{app:collab}}\begin{sloppypar}\hyphenpenalty=5000\widowpenalty=500\clubpenalty=5000\input{B2G-15-008-authorlist.tex}\end{sloppypar}
\end{document}

%% file: B2G-15-008-authorlist.tex
\textbf{Yerevan Physics Institute,  Yerevan,  Armenia}\\*[0pt]
V.~Khachatryan, A.M.~Sirunyan, A.~Tumasyan
\vskip\cmsinstskip
\textbf{Institut f\"{u}r Hochenergiephysik,  Wien,  Austria}\\*[0pt]
W.~Adam, E.~Asilar, T.~Bergauer, J.~Brandstetter, E.~Brondolin, M.~Dragicevic, J.~Er\"{o}, M.~Flechl, M.~Friedl, R.~Fr\"{u}hwirth\cmsAuthorMark{1}, V.M.~Ghete, C.~Hartl, N.~H\"{o}rmann, J.~Hrubec, M.~Jeitler\cmsAuthorMark{1}, A.~K\"{o}nig, I.~Kr\"{a}tschmer, D.~Liko, T.~Matsushita, I.~Mikulec, D.~Rabady, N.~Rad, B.~Rahbaran, H.~Rohringer, J.~Schieck\cmsAuthorMark{1}, J.~Strauss, W.~Waltenberger, C.-E.~Wulz\cmsAuthorMark{1}
% ...
\vskip\cmsinstskip
% ...
1:~~Also at Vienna University of Technology, Vienna, Austria\\
% ...
